\documentclass[letterpaper]{article} % DO NOT CHANGE THIS
\usepackage{sty/aaai2026}  % DO NOT CHANGE THIS
\usepackage{times}  % DO NOT CHANGE THIS
\usepackage{helvet}  % DO NOT CHANGE THIS
\usepackage{courier}  % DO NOT CHANGE THIS
\usepackage[hyphens]{url}  % DO NOT CHANGE THIS
\usepackage{graphicx} % DO NOT CHANGE THIS
\usepackage{natbib}  % DO NOT CHANGE THIS AND DO NOT ADD ANY OPTIONS TO IT
\usepackage{caption} % DO NOT CHANGE THIS AND DO NOT ADD ANY OPTIONS TO IT

\usepackage{multirow}

\usepackage[T1]{fontenc}
\usepackage{ifthen}
\usepackage{algorithm}
\usepackage[noEnd=true,indLines=true,commentColor=black]{sty/algpseudocodex}
\usepackage{amsmath,amssymb,amsthm}
\usepackage{booktabs}
\usepackage{cleveref}
\usepackage{siunitx}
\usetikzlibrary{calc,math}
\usepackage{sty/custom}

\usepackage{newfloat}
\usepackage{listings}
\DeclareCaptionStyle{ruled}{labelfont=normalfont,labelsep=colon,strut=off} % DO NOT CHANGE THIS
\floatstyle{ruled}
\newfloat{listing}{tb}{lst}{}
\floatname{listing}{Listing}
\title{Local Guidance for Configuration-Based Multi-Agent Pathfinding}
\author{Tomoki Arita$^{1,2}$, Keisuke Okumura$^{1,3}$}
\affiliations{
  $^1$National Institute of Advanced Industrial Science and Technology (AIST), Japan
  \\
  $^2$Keio University, Japan
  \\
  $^3$University of Cambridge, UK
  \\
  \{arita.tomoki, okumura.k\}@aist.go.jp
}

\begin{document}

\nocopyright
\maketitle

\begin{abstract}
\emph{Guidance} is an emerging concept that improves the empirical performance of real-time, sub-optimal multi-agent pathfinding (MAPF) methods.
It offers additional information to MAPF algorithms to mitigate congestion on a \emph{global} scale by considering the collective behavior of all agents across the entire workspace.
This global perspective helps reduce agents' waiting times, thereby improving overall coordination efficiency.
In contrast, this study explores an alternative approach: providing \emph{local} guidance in the vicinity of each agent.
While such localized methods involve recomputation as agents move and may appear computationally demanding, we empirically demonstrate that supplying informative spatiotemporal cues to the planner can significantly improve solution quality without exceeding a moderate time budget.
When applied to LaCAM, a leading configuration-based solver, this form of guidance establishes a new performance frontier for MAPF.
\end{abstract}

\section{Introduction}
A holy grail challenge in multi-agent pathfinding (MAPF) research is to develop scalable yet real-time methods for computing collision-free paths with reasonable solution quality.
Given the computational intractability of finding optimal solutions~\cite{yu2013structure,banfi2017intractability}, the field has increasingly shifted its focus toward efficient suboptimal approaches.
Indeed, the past decade has seen the emergence of several powerful algorithms capable of handling hundreds of agents.
Notable examples include \emph{PIBT}~\cite{okumura2022priority} and \emph{LaCAM}~\cite{okumura2023lacam,okumura2023lacam2}, which rapidly and reliably deliver feasible solutions, even in densely populated scenarios where other MAPF algorithms falter.

The evolution of these suboptimal algorithms reveals a new MAPF challenge, particularly in dense setups: planners must now mitigate \emph{congestion} in addition to deriving collision-free motions.
As one might intuitively expect, when many agents gather in a specific spatiotemporal region, most agents are forced to wait in place, wasting time, and thus significantly degrading the planning performance.

This awareness leads to a series of studies introducing \emph{guidance}~\cite{zhang2024guidance}.
The guidance itself does not derive collision-free paths;
rather, it supports other heuristic search methods by providing congestion mitigation information, as like the left-hand side rule in traffic management.
Such guidance indeed plays a critical role in achieving high-performance MAPF systems, as we see examples in winning strategies in the MAPF competition~\cite{chan2024league,jiang2024scaling,yukhnevich2025enhancing}.
The exact form of guidance varies from study to study, but it is typically constructed through a \emph{global} consideration of the entire environment and the entire team of agents~\cite{han2022optimizing,chen2024traffic,kato2025congestion}.
Such global guidance also discards time-parametrized information about agent trajectories to ensure they remain trackable.

This study, in contrast, develops the notion of \emph{local} guidance around individual agents.
The underlying rationale is that global guidance, at one extreme, can be viewed as a coarse solution concept that relaxes the strict collision constraints of exact solutions, which lie at the other extreme.
Local guidance is designed to lie between these two ends of the spectrum:
it is more informative than global guidance, incorporating spatiotemporal awareness, yet less computationally demanding than computing fully collision-free, (near-)optimal paths.
As a result, it offers a practical compromise that balances planning effort and solution quality.

We instantiate the concept of local guidance using a leading suboptimal MAPF algorithm, LaCAM, which relies on a configuration---the simultaneous positions of all agents---as its planning representation.
Empirical results show that this enhancement significantly improves solution quality, outperforming its global counterpart while maintaining real-time responsiveness, typically within a few seconds even for $1,000$ agents.
In the most extreme case, we observe a 50\% reduction in solution cost compared to the original LaCAM.
Moreover, the local guidance enhancement mostly surpasses the state-of-the-art anytime MAPF solver~\cite{okumura2024lacam3}.
Together, these results further push the frontier of real-time MAPF, promoting the practicality of MAPF in real-world applications.
The code is publicly available at \url{https://github.com/allegorywrite/lg_lacam}.

\section{Preliminaries and Related Work}

\paragraph{Problem Definition.}
This paper focuses on one-shot, classical MAPF formulation~\cite{stern2019def}.
The system consists of a set of agents $A = \{1, 2, \ldots, n\}$ and a graph $G=(V, E)$.
All agents act synchronously according to a discrete wall-clock time.
At each timestep, each agent can stay in place or move to an adjacent vertex.
Vertex and edge collisions are considered, i.e., two agents cannot occupy the same vertex simultaneously, and two agents cannot swap their occupied vertices within one timestep move.
Then, we aim to find a finite action sequence for each agent $i \in A$ to its assigned goal $g_i \in V$ from a given start $s_i \in V$.
The solution quality is assessed by \emph{flowtime} (aka. sum-of-costs; SoC), which sums the travel time of each agent until it stops at the target location and no longer moves.

\paragraph{LaCAM} \emph{(lazy constraints addition search)}~\cite{okumura2023lacam} is a search-based algorithm that finds unbounded suboptimal solutions quickly, leading to large-scale MAPF studies due to its scalability and real-time responsiveness~\cite{shen2023tracking}.
Although its planning ability to handle densely populated, complicated instances at scale is remarkable~\cite{okumura2023lacam2}, the resultant solutions are known to be highly suboptimal~\cite{chan2024anytime,tan2025reevaluation}.
This gives rise to two practical challenges: \emph{(i)}~finding better initial solutions with an acceptable computational overhead, or \emph{(ii)}~accelerating the refinement process to reach optimal solutions from feasible initial ones~\cite{okumura2024lacam3}.
This study focuses on the first challenge, but we will later demonstrate that improving the initial solutions leads to better final outcomes under the severe time limit.

\paragraph{How LaCAM Works.}
Let a \emph{configuration} $\Q \in V^n$ be the locations for all agents.
For instance, the start configuration is $\mathcal{S} = (s_1, s_2, \ldots, s_n)$ and the goal is $\mathcal{G} = (g_1, g_2, \ldots, g_n)$.
Then, MAPF is reduced to single-agent pathfinding from $\mathcal{S}$ to $\mathcal{G}$ on a graph consisting of configurations.
Two configurations are neighboring if any agent can move from one to the other within one timestep, and there are no collisions during this transition.
This connectivity defines a graph structure;
thus, any search algorithm, such as depth-first search (DFS), can derive a solution to MAPF.
Within this scheme, LaCAM is a variant of DFS, but with a lazy successor generation that avoids inspecting all successor configurations at once, which is exponential to the number of agents.
This generation process utilizes another MAPF algorithm, termed \emph{configuration generator}, to guide the search towards the goal $\mathcal{G}$ efficiently.

\paragraph{Configuration Generator} generates a connected configuration $\Q_{k+1}$ given a configuration $\Q_k$.
LaCAM's performance in terms of both speed and quality is heavily influenced by this process as it determines the search direction.
Typically, PIBT is employed to prioritize speed.

\paragraph{PIBT} \emph{(priority inheritance with backtracking)}~\cite{okumura2022priority} is a computationally lightweight function that synthesizes a neighboring configuration given another \Q.
As fast configuration generation is central to various MAPF situations, PIBT has gained notable popularity in recent years.
For each agent $i$, it uses a sorted list of candidate actions $v$ from $\Q[i]$, termed \emph{preference}, where $v \in \neigh(\Q[i]) \cup \{ \Q[i] \}$ and \neigh denotes a set of adjacent vertices.

\paragraph{Preference Construction in PIBT} originally employs sorting actions in a lexicographic and ascending order with $\langle \dist(v, g_i), \epsilon\rangle$, where $\dist$ denotes the shortest path distance on $G$ and $\epsilon$ is a random number to break ties.
Although this certainly guides agents towards their respective goals, it can also result in greedy, myopic behaviors, which are often accompanied by deadlock, livelock, or congestion.
As PIBT gets popularity, numerous methods have been developed to optimize the preference: Monte-Carlo sampling~\cite{okumura2024lacam3}, imitation learning~\cite{jiang2025deploying,jain2025lagat}, DFS~\cite{gandotra2025anytime}, and tiebreaking~\cite{okumura2025lightweight}, among others.

\paragraph{Guidance} is one of these attempts in PIBT's preference optimization, which encourages agents to follow the congestion mitigation heuristic to optimize the traffic flow \emph{globally}.
Various approaches exist to construct guidance, such as handcrafted~\cite{wang2008fast,yukhnevich2025enhancing}, heuristic search~\cite{han2022optimizing,chen2024traffic,kato2025congestion}, or blackbox optimization~\cite{zhang2024guidance}.
To improve LaCAM, we are also interested in better preferences with guidance,  but \emph{locally}, i.e., mitigating congestion surrounding each agent in each configuration.
\Cref{fig:concept} sketches this difference between global and local ones and how to integrate them within the search.
The remainder of this paper crystallizes this notion with empirical evidence.

{
  \begin{figure}[t!]
    \centering
    \includegraphics[width=0.9\linewidth]{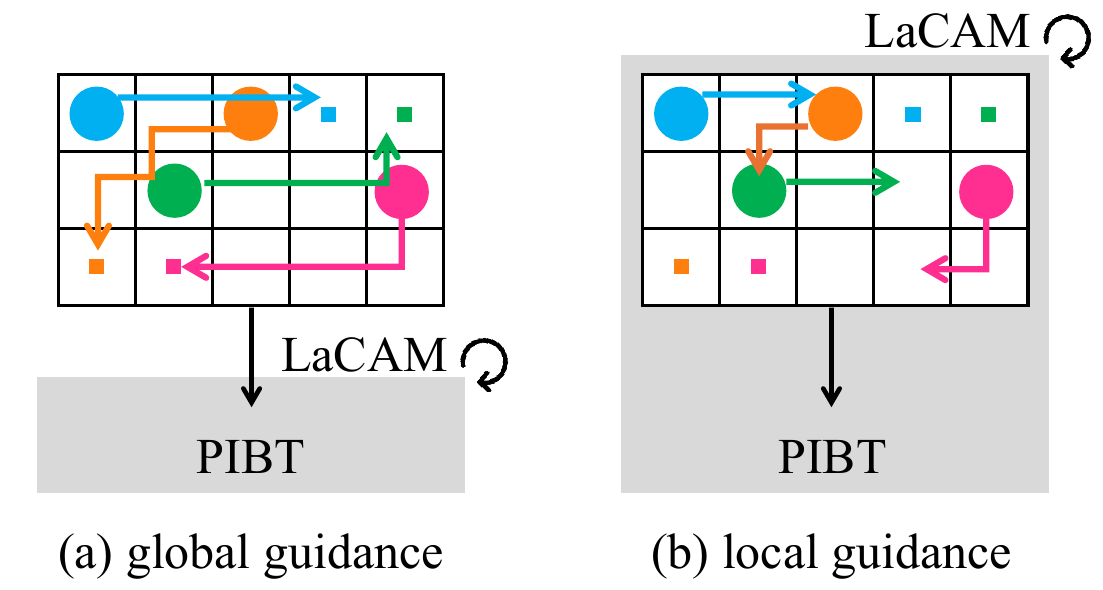}
\caption{
Concept of global and local guidance with LaCAM.
Goals for agents are marked with colored boxes.
The gray-shaded regions correspond to configuration generation.
}
    \label{fig:concept}
  \end{figure}
}

\section{Method}
This study follows the guidance construction based on heuristic search~\cite{han2022optimizing,chen2024traffic,okumura2024lacam3}, which is known to effectively improve representative MAPF algorithms, including LaCAM, without the need for prior preparation (e.g., offline data collection).
Specifically, we utilize \cref{algo:guidance} to construct agent-wise paths $\Phi$ starting from a given configuration \Q towards the goal configuration \G.
\cref{algo:guidance} adopts \emph{windowed} planning, where the planning horizon is limited by $w$ timesteps ahead, which is known to be a technique for attaining computational efficiency~\cite{silver2005cooperative,li2021lifelong}.
In the pseudocode, $\{c, c_T\}: V \times V \mapsto \mathbb{R}$ represent the cost functions;
$c$ is for the stage cost while $c_T$ is for the terminal cost.

{
\begin{algorithm}[th!]
\caption{Guidance Construction}
\label{algo:guidance}
\begin{algorithmic}[1]
\small
\Input{configuration $\Q \in V^n$, goals $\G = (g_1, g_2, \ldots, g_n) \in V^n$}
\Output{$\Phi \in V^{n\times (w + 1)}$\quad\textbf{params:}~$w \in \mathbb{N}_{>0}$}
\State initialize $\Phi$
\label{algo:guidance:init}
\While{termination condition not met}
\label{algo:guidance:termination}
\For{$i \in A$}
\label{algo:guidance:order}
\State $\begin{aligned}
\Phi[i] \leftarrow& \argmin_{\pi \in \Pi}
\sum_{t = 0}^{w-1} c(\pi[t], \pi[t+1]) + c_T(\pi[w], g_i)
\\
&\Pi: \text{paths}~\text{from}~Q[i]~\text{on}~G~\text{with length}~w+1
\end{aligned}$
\label{algo:guidance:pathfinding}
\EndFor
\label{algo:guidance:for-end}
\EndWhile
\State \Return $\Phi$
\end{algorithmic}
\end{algorithm}
}

Observe that the structure abstracted in \cref{algo:guidance} captures various MAPF techniques by adjusting the cost definition.
For instance, seminal prioritized planning (PP)~\cite{erdmann1987multiple,silver2005cooperative} is considered to be a type of \cref{algo:guidance} with an infinite window size and hard constraints for goals and collisions that must never be violated.
These hard constraints are represented by infinite costs when they are not met.
LNS2~(large neighborhood search)~\cite{li2022mapf} is another instantiation of \cref{algo:guidance}, which starts from collision-containing infeasible paths and gradually resolves them by repairing paths for a subset of the entire team.
LNS2 uses soft collision constraints, which allow for collisions during replanning but are negatively incentivised.

We leverage this widely-used algorithmic structure to construct the local guidance.
\emph{It should mitigate congestion locally while remaining computationally efficient}, given that it is invoked at each search iteration of LaCAM.
For this purpose, we employ windowed planning, where $w$ typically ranges from 5 to 20.
This approach naturally focuses on each agent's local situation while avoiding excessive computational overhead.
Congestion mitigation is achieved through carefully designed soft collision constraints.
In what follows, we first describe how to exploit the guidance paths, and then specify each component of \cref{algo:guidance}.

\paragraph{Exploitation.}
Once the guidance paths $\Phi$ have been obtained, a design choice must be made regarding how to incorporate them into PIBT.
We follow the latest LaCAM implementation~\cite{okumura2024lacam3}, which uses the global guidance.
Specifically, PIBT uses a scoring
\begin{align}
\langle \funcname{Ind}\bigl[\Phi[i][1] \neq v\bigr], \dist(v, g_i), \epsilon \rangle
\label{eq:pref}
\end{align}
for the preference construction, where \funcname{Ind} is an indicator function that returns one only if the condition is true, otherwise zero.
This encourages agents to follow the guidance paths while minimizing the computational overhead required to extract the guidance information.
Note that this preserves LaCAM's completeness guarantee for MAPF.

\paragraph{Cost and Path Planning (\cref{algo:guidance:pathfinding}).}
The cost design is core for achieving effective guidance.
Inspired from the studies on path-based global guidance~\cite{han2022optimizing,chen2024traffic,okumura2024lacam3}, we use a lexicographic cost:
\begin{align}
  c(\pi[t], \pi[t+1]) &:= \langle 1 + \alpha\cdot\funcname{Ind}[\chi > 0], \chi \rangle,
  \label{eq:cost}
\\
c_T(\pi[w], g_i) &:= \langle \dist(\pi[w], g_i), 0 \rangle,
\end{align}
where $\chi$ denotes the number of collisions for $(\pi[t], \pi[t+1])$ with other paths in $\Phi$.
The hyperparameter $\alpha \in \mathbb{R}_{\geq 0}$ determines the importance of collisions;
our implementation sets $\alpha=3$ by default, based on the pilot study.
For agent $i$ at configuration \Q, this will find the first fragment of the path from $\Q[i]$ to $g_i$ with the fewer collisions within the window, while avoiding overly conservative motion.
Such a path can be quickly identified by space-time \astar~\cite{silver2005cooperative}.
This is because the search space is limited by windowed planning, which enables the memory needed for \astar to be pre-allocated.%
\footnote{
An alternative to space-time \astar is SIPP~\cite{phillips2011sipp}, a fast single-agent pathfinding method for dynamic environments.
We tested SIPP, but did not achieve its speed benefits over space-time \astar.
}

\paragraph{Plan Iteration (\cref{algo:guidance:termination}).}
As \cref{algo:guidance} plans paths sequentially, the first planned agent is less constrained by others in terms of collision penalty, while the last planned agent is heavily constrained.
To resolve this unfairness, we iterate the path planning at Lines~\ref{algo:guidance:order}--\ref{algo:guidance:for-end} several times.
Empirically, we found that a few iterations, e.g., twice, are sufficient to obtain effective guidance.
Meanwhile, repeated planning for the entire team adds non-negligible computational overhead.
Therefore, as explained next, if available, we reuse the previous guidance paths to approximate the first iteration.

\paragraph{Initialization (\cref{algo:guidance:init}).}
As LaCAM adopts DFS, the search within LaCAM explores configurations in a connected sequence.
For example, if the current search iteration generates $\Q_{k+1}$ from $\Q_k$, the previous iteration should be for $\Q_{k-1}$, which is connected to $\Q_k$.
Utilizing this structure, we initialize the guidance $\Phi_k$ for $\Q_{k}$ by the previous result $\Phi_{k-1}$ for $\Q_{k-1}$ with \cref{algo:init}, at \cref{algo:guidance:init} in \cref{algo:guidance}.
This procedure shifts the guidance path for agent $i$ by one timestep if $i$ is at the intended location of the previous guidance.
This initialization allows us to omit the plan iteration, which results in significant time savings.

{
\begin{algorithm}[th!]
\caption{Initialization}
\label{algo:init}
\begin{algorithmic}[1]
\small
\Input{configuration $\Q_k$, previous guidance $\Phi_{k-1}$}
\Output{$\Phi_k$ (each path is initialized as empty)}
\For{$i \in A$}
\If{$\Q_k[i] = \Phi_{k-1}[i][1]$}
\For{$t = 0, \ldots, w - 1$}
\State $\Phi_k[i][t] \leftarrow \Phi_{k-1}[i][t+1]$
\EndFor
\EndIf
\EndFor
\State \Return $\Phi_k$
\end{algorithmic}
\end{algorithm}
}

{
\setlength{\tabcolsep}{1pt}
\newcommand{\entry}[4]{
\begin{scope}[xshift=0.235*#3\linewidth]
{
\node[anchor=south west] at (0, 0)
{\includegraphics[width=0.24\linewidth]{fig/raw/heatmaps/maze-128-128-10_#2_visits_heatmap}};
\node[] at (2.3, 4.85) {\small #1};
\node[] at (2.3, 4.55) {\scriptsize cost improvement: #4};
%
% local guidance effective
\draw[blue,line width=2pt,opacity=1] (0.25, 2.1)
-- node[pos=0.5,below,fill,circle,text=white,inner sep=1pt] {\small B}
++(1.2, 0) -- ++(0, 0.85) -- ++(-1.2, 0) -- cycle;
%
% global guidance effective
\draw[cyan,line width=2pt] (0.4, 3.15) -- ++(1.7, 0) -- ++(0, 0.55)
-- node[pos=0.5,above,fill,circle,text=white,inner sep=1pt] {\small A}
++(-1.7, 0) -- cycle;
%
% usage
\node[anchor=north west] at (0.2, 0.1)
{\includegraphics[width=0.225\linewidth,clip,trim={0.9cm 0.9cm 0 0}]{fig/raw/heatmaps/maze-128-128-10_#2_visits_histogram}};
\node[anchor=north west,rotate=0] at (0.1, 0.2) {\tiny freq.};
\node[anchor=north west] at (1.5, -0.92) {\tiny number of visits};
}
\end{scope}
}
\begin{figure*}[tp!]
\centering
\begin{tikzpicture}
\entry{LaCAM}{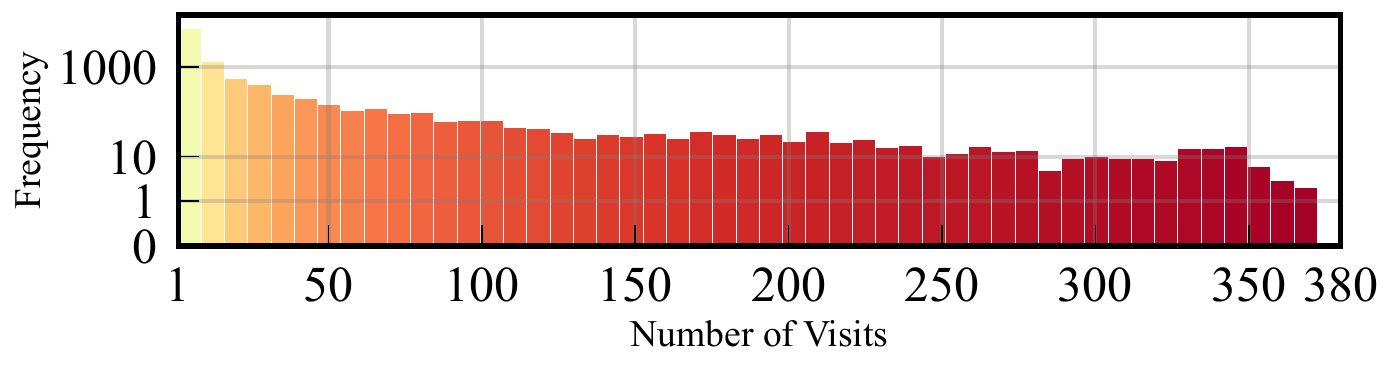}{0}{0.0\%}  % 351278  -> tikz math cannot calculate this...
\entry{GG}{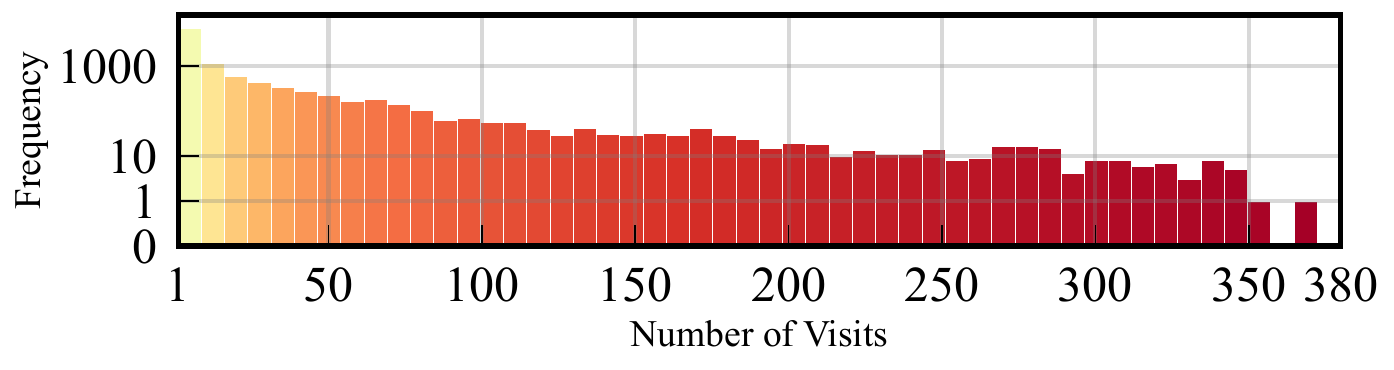}{1}{16.9\%}          % (351278 - 291977) / 351278
\entry{LG}{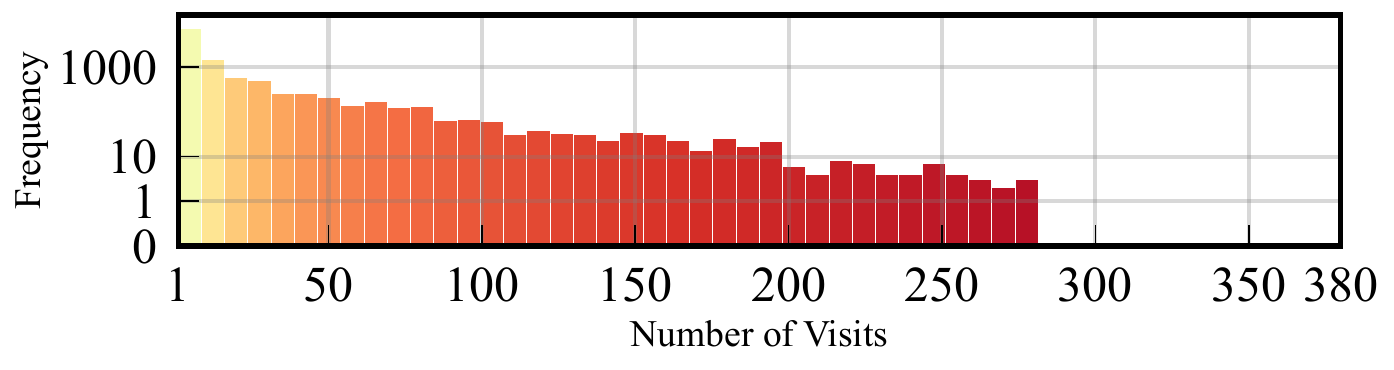}{2}{38.1\%}                % (351278 - 217579) / 351278
\entry{GG+LG}{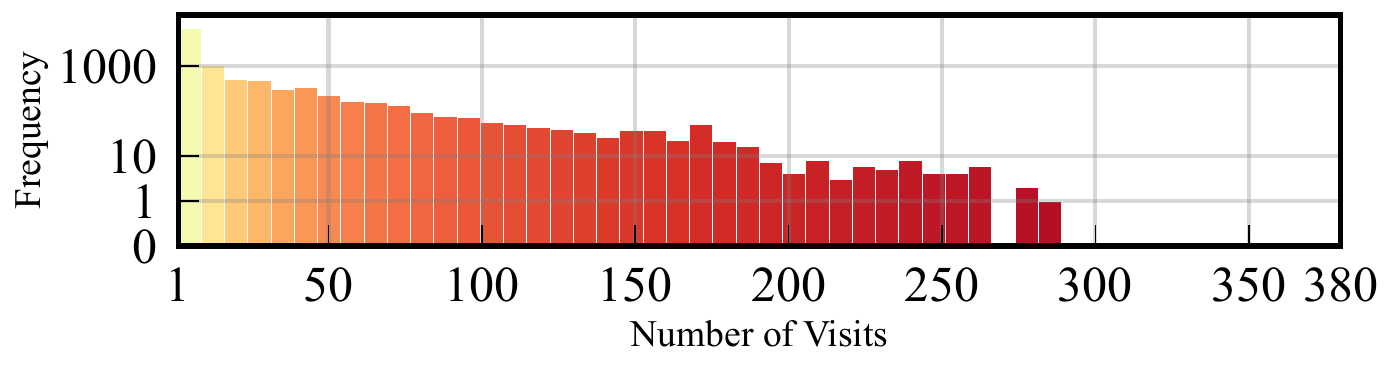}{3}{38.4\%}         % (351278 - 216559) / 351278
\node[anchor=south west] at (0.95\linewidth, -0.1)
{\includegraphics[width=0.043\linewidth,clip,trim={0 0 0.8cm 0}]{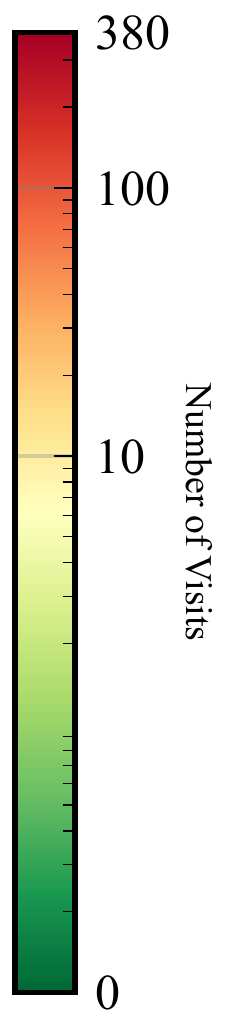}};
\node[anchor=south east,rotate=-90] at (0.98\linewidth, 0.15) {\small number of visits};
\end{tikzpicture}
\caption{Vertex usage in LaCAM solutions, where congestion is implied by warm colors.
The same start-goal instance comprising 1,000 agents on \mapname{maze-128-128-10} is used across different solvers.
The cost improvement percentage is shown at the top, with LaCAM as the baseline.
The bottom shows the distribution among 14,818 vertices of how many times each vertex is used by any agent.
In the middle, area-A shows the effect of global guidance, while area-B for local guidance.
}
\label{fig:heatmap}
\end{figure*}
}

\paragraph{Agent Order (\cref{algo:guidance:order}).}
Similar to PP, whose performance can vary depending on the order of the agents~\cite{wu2020multi}, the agent order at \cref{algo:guidance:order} in \cref{algo:guidance} also affects the resultant guidance quality, especially when the number of plan iterations is limited.
For the construction of guidance $\Phi_k$, we sort the agent list $A$ in descending order according to the number of collisions in the previous guidance $\Phi_{k-1}$.
The rationale is to relax the unfairness between agents, as discussed in the plan iteration.

\paragraph{Global Guidance Consideration.}
It is possible to incorporate global guidance into local guidance construction to mitigate congestion on a global scale.
Assume that the global guidance $\Psi$ is provided, and it takes the form of agent-wise paths similar to our local guidance, i.e., $\Psi[i] \in V^m$, as seen in~\cite{han2022optimizing,chen2024traffic,okumura2024lacam3}.
Then, we modify the cost function for agent $i$, \cref{eq:cost}, to reflect the global guidance as
\begin{align}
c(\pi[t], \pi[t+1]) := \langle 1 + \alpha\cdot\funcname{Ind}[\chi > 0], \delta(\pi[t+1]), \chi \rangle.
\label{eq:pref2}
\end{align}
The new interim term $\delta(v)$ measures the distance from $\Psi[i]$, which is computed by lazy breadth-first search upon query.
The terminal cost, $c_T(\cdot)$, remains unchanged except for a dimension match.
The updated cost design incentivizes agents to adhere to global guidance while prioritizing the mitigation of local congestion.

\paragraph{Integration with Swap.}
The \emph{swap} technique is an important technique for stabilizing LaCAM's planning performance~\cite{okumura2023lacam2}.
It reverses the PIBT preference for specific agents, enabling those who need to exchange locations within narrow corridors to escape local deadlocks.
Local guidance can co-exist with the swap technique:
if the swap situation is identified for an agent, we simply discard the guidance term in \cref{eq:pref} and use the reverse scoring $\langle 0, -\dist(v,g_i), \epsilon\rangle$.

\paragraph{Time Complexity of Guidance Construction.}
Running $w$-windowed space-time A* on a graph $G = (V, E)$ has a time complexity of $O(w|E| + w|V|\log (w|V|))$, given that Dijkstra’s algorithm on the time-expanded graph of $G$ with horizon $w$ exhibits the same order.
The local guidance construction executes this process for $n$ agents, $m$ times, resulting in an overhead of $O(mn(w|E| + w|V|\log (w|V|)))$.
As demonstrated in the experiments, $m=1$ is typically sufficient for solution improvement, with the plan initialization (\cref{algo:init}) operating in $O(nw)$.
Assuming $G$ is a four-connected grid, typically used in MAPF benchmarks where $O(|E|)=O(|V|)$, the practical overhead for local guidance construction is thus derived as $O(nw|V|\log (w|V|))$.

\section{Evaluation}
Experiments were conducted on a laptop equipped with Intel Ultra 9 185H and \SI{62}{\giga\byte} of RAM.
We use ``random'' scenarios from the MAPF benchmark~\cite{stern2019def}, consisting of 25 instances for each map and each number of agents (up to $1,000$), which is commonly used to evaluate MAPF methods.
Unless specified, the planner's time limit is set to \SI{30}{\second}, following~\cite{okumura2023lacam2};
otherwise, the planner is considered unsuccessful in solving the instance.
This section compares the following methods:
\begin{itemize}
\item \textbf{LaCAM} denotes a basic implementation without any guidance terms, aligned with~\cite{okumura2023lacam2}.
\item \textbf{GG} uses global guidance in \cref{eq:pref}.
  The guidance is constructed with \emph{space utilization optimization (SUO)}, as described in~\cite{han2022optimizing,okumura2024lacam3}.\footnote{
  We also tested the traffic flow-based method~\cite{chen2024traffic}, but found that SUO with LaCAM produced superior results.
  }
  This method precomputes time-independent, less-congested paths for each agent, which reduces the effort required for collision resolution during the planning stage.
\item \textbf{LG} uses the local guidance in \cref{eq:pref}.
  Unless noted, LG reuses the previous guidance information and refines it only once.
  The window size is set to $w=20$, and the collision penalty $\alpha=3$.
\item \textbf{LG+GG} uses both global and local terms with \cref{eq:pref2}.
\item \textbf{LNS2}~\cite{li2022mapf} is another leading unbounded suboptimal MAPF algorithm, which falls into the same category as LaCAM for the solution quality guarantee.
\end{itemize}
All LaCAM-based methods use the swap technique.

{
\setlength{\tabcolsep}{1pt}
\newcommand{\entry}[4]{
\begin{minipage}{0.19\linewidth}
\centering
  \setlength{\baselineskip}{0.75\baselineskip}
  \begin{tikzpicture}
  \scriptsize
  \node[anchor=south] at (-1.6, 0.15) {\mapname{#1}};
  \node[anchor=south] at (-1.6, -0.15) {#2~(#3)};
  \node[anchor=north east] at (0, 0)
  {\includegraphics[width=1\linewidth,height=1\linewidth]{fig/raw/main_results/time_flowtime_#1}};
  \node[anchor=north east] at (-0.15, -0.15)
  {\includegraphics[width=0.25\linewidth]{fig/raw/maps/#1}};
  \ifthenelse{\equal{#4}{1}}{
  \begin{scope}[xshift=-1.0cm,yshift=-3.0cm]
  \draw[] (0, 0) --
  node[pos=0]{\small $\times$}
  node[pos=0.25]{\small $\bullet$}
  node[pos=0.5]{$\blacktriangle$}
  node[pos=0.75]{\tiny $\blacksquare$}
  node[pos=1.0]{\tiny $\bigstar$}
  node[right,pos=0]{\tiny 200}
  node[right,pos=0.25]{\tiny 400}
  node[right,pos=0.5]{\tiny 600}
  node[right,pos=0.75]{\tiny 800}
  node[right,pos=1]{\tiny 1000}
  node[above right=0.2,pos=1]{\tiny agents}
  (0.2, 0.8);
  \end{scope}
  }{}

  \end{tikzpicture}
  \end{minipage}
}
\newcommand{\ylabel}{\rotatebox{90}{\small \hspace{-4.0em}{$\leftarrow$ flowtime / LB}}}
\begin{figure*}[t!]
  \centering

  \begin{tabular}{cc}
    \rotatebox{90}{\hspace{0.5em}\small $\leftarrow$ cost reduction} &
    \includegraphics[width=0.97\linewidth]{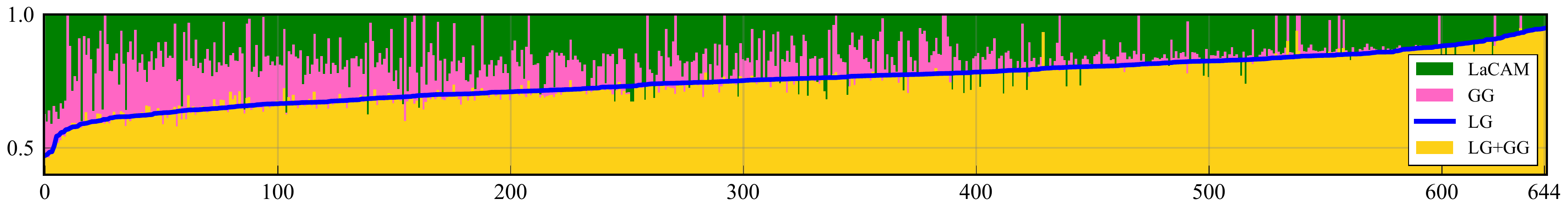} \\
    \rotatebox{90}{\hspace{0.8em}\small $\leftarrow$ runtime [\SI{}{\second}]} &
    \includegraphics[width=0.97\linewidth]{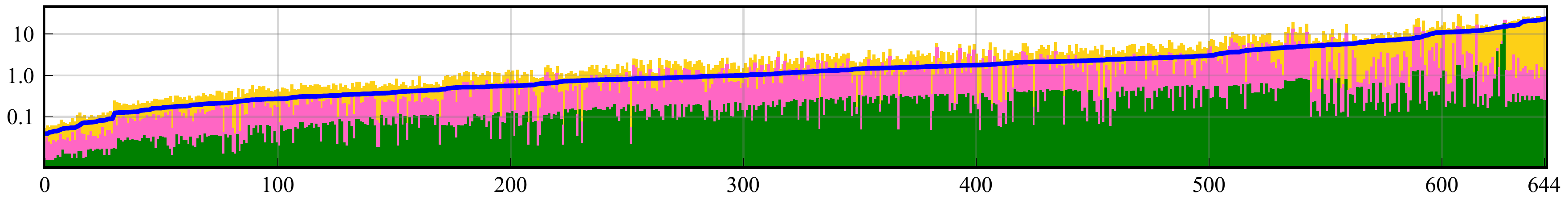} \\
    & {\small instances}
  \end{tabular}
  \medskip
  \begin{tabular}{cccccccccc}
    \ylabel &
    \entry{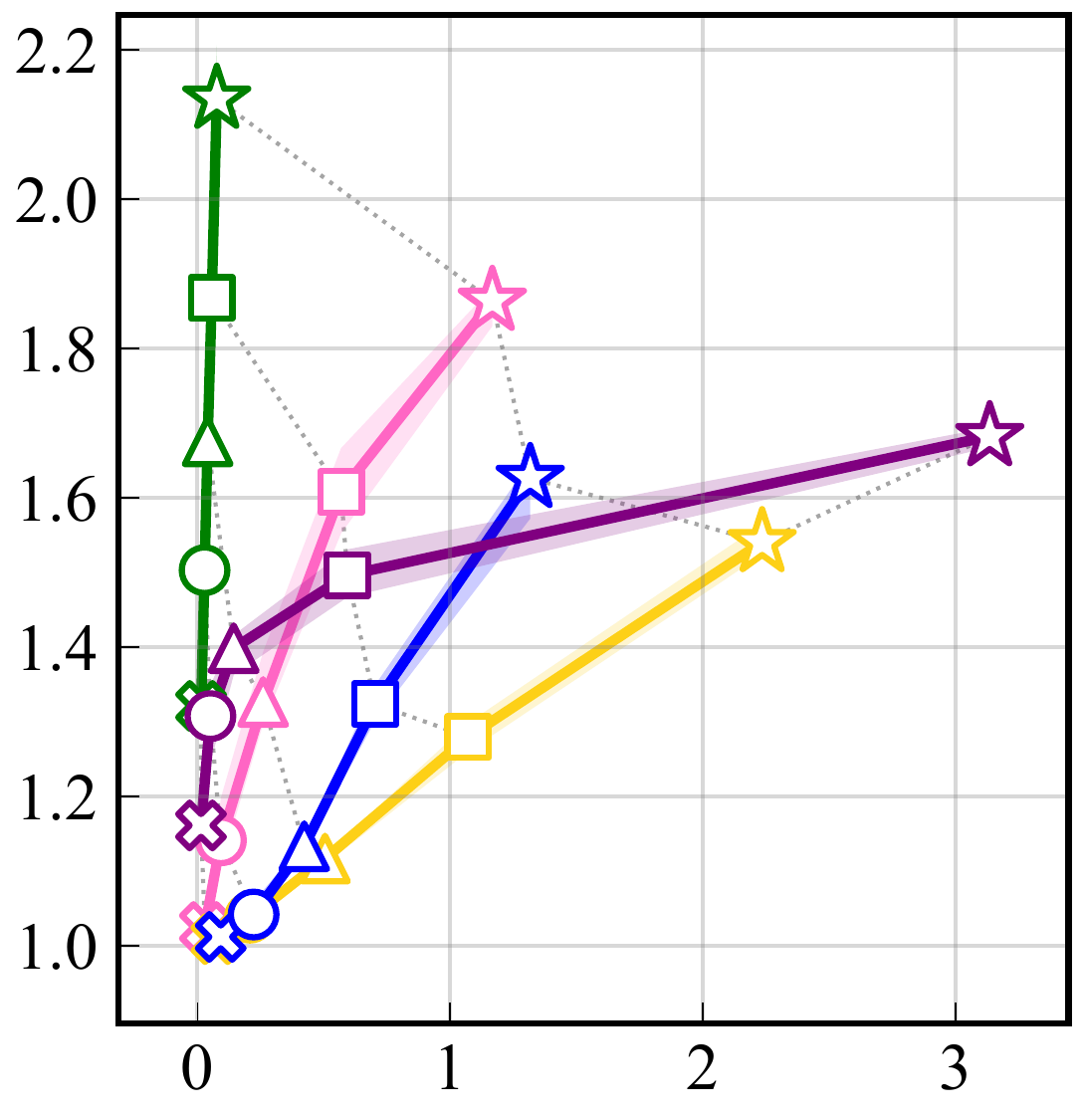}{48x48}{2,304}{1}&
    \entry{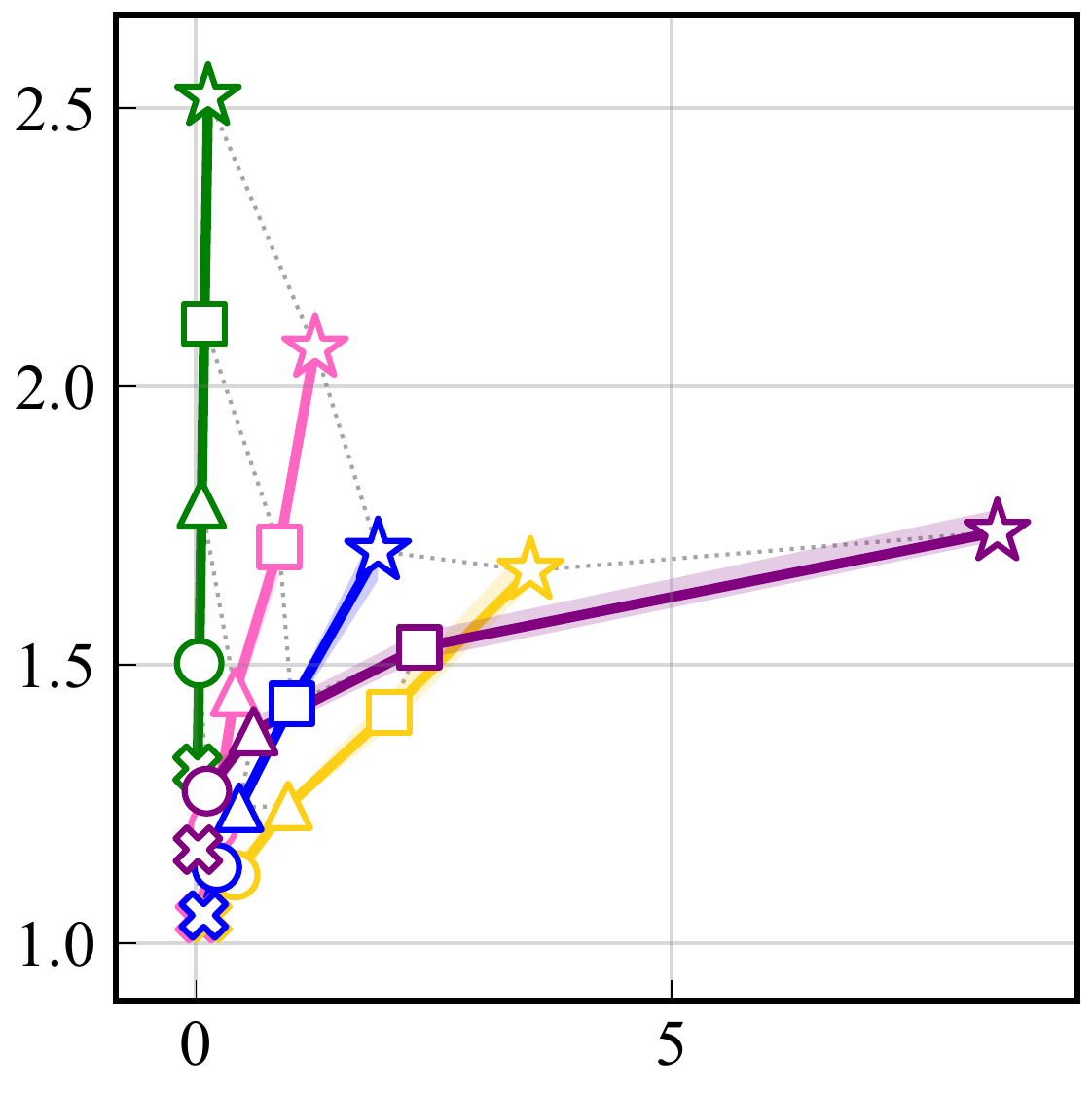}{64x64}{3,687}{}&
    \entry{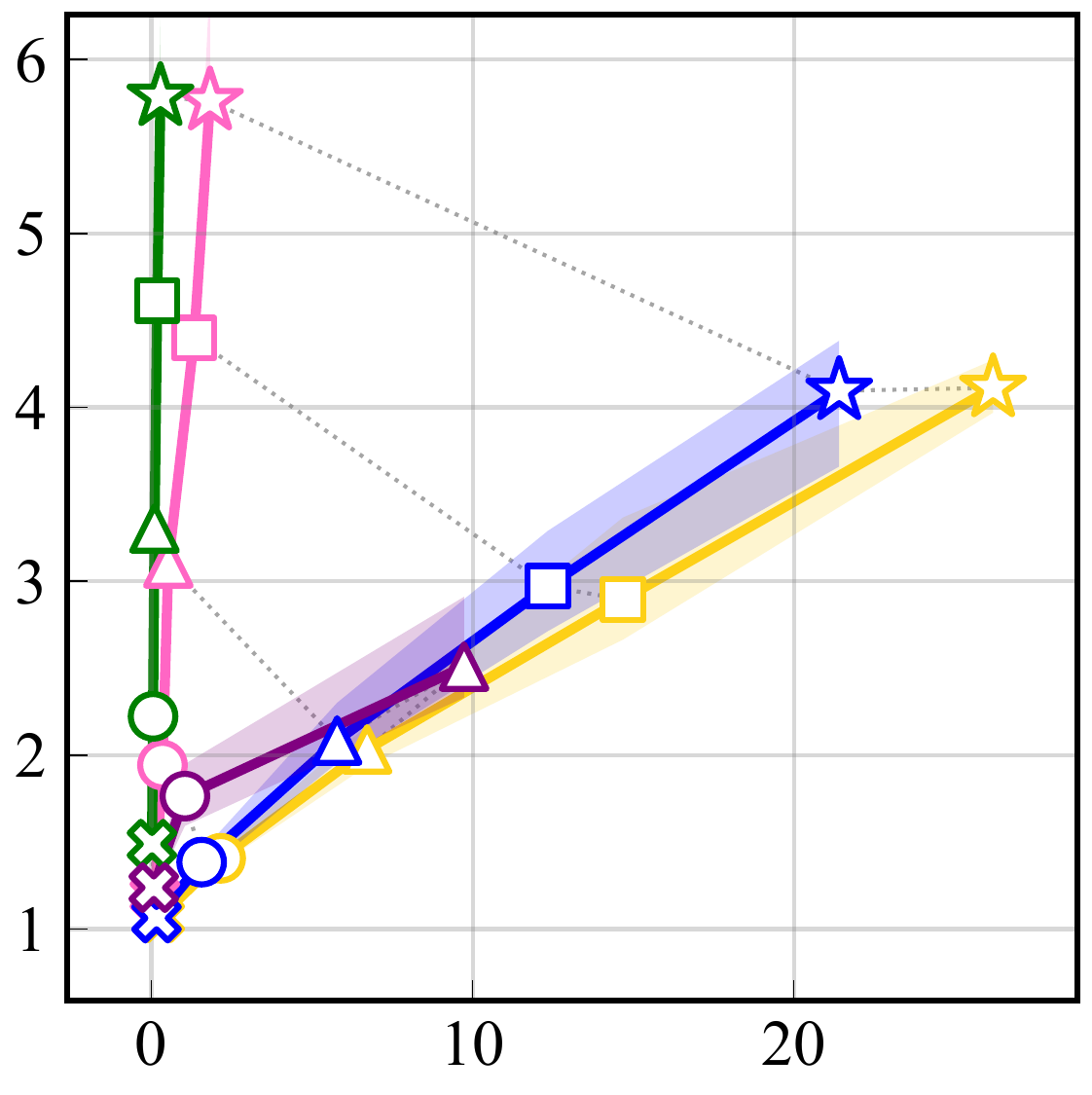}{65x81}{2,445}{}&
    \entry{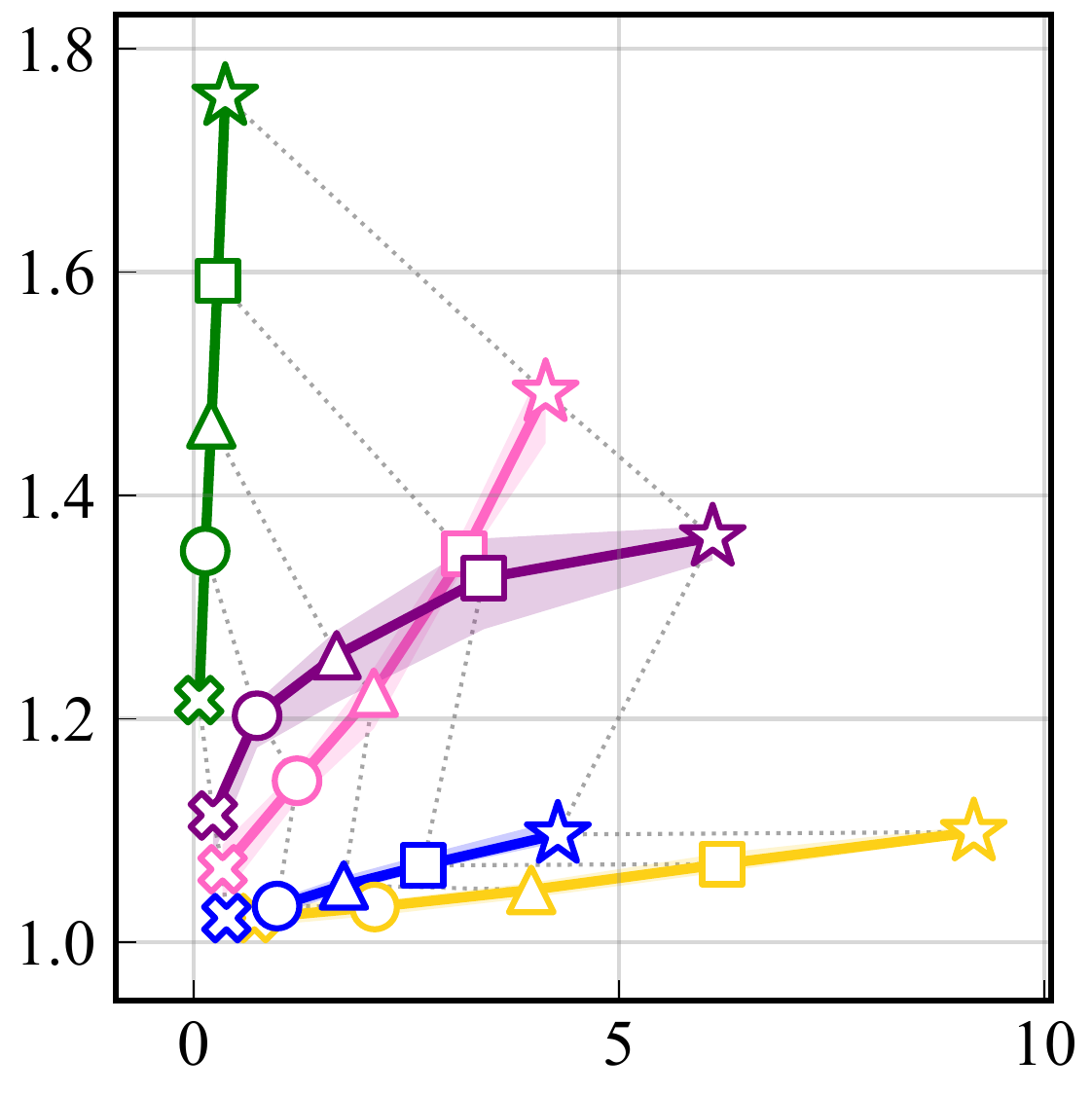}{128x128}{14,818}{}&
    \entry{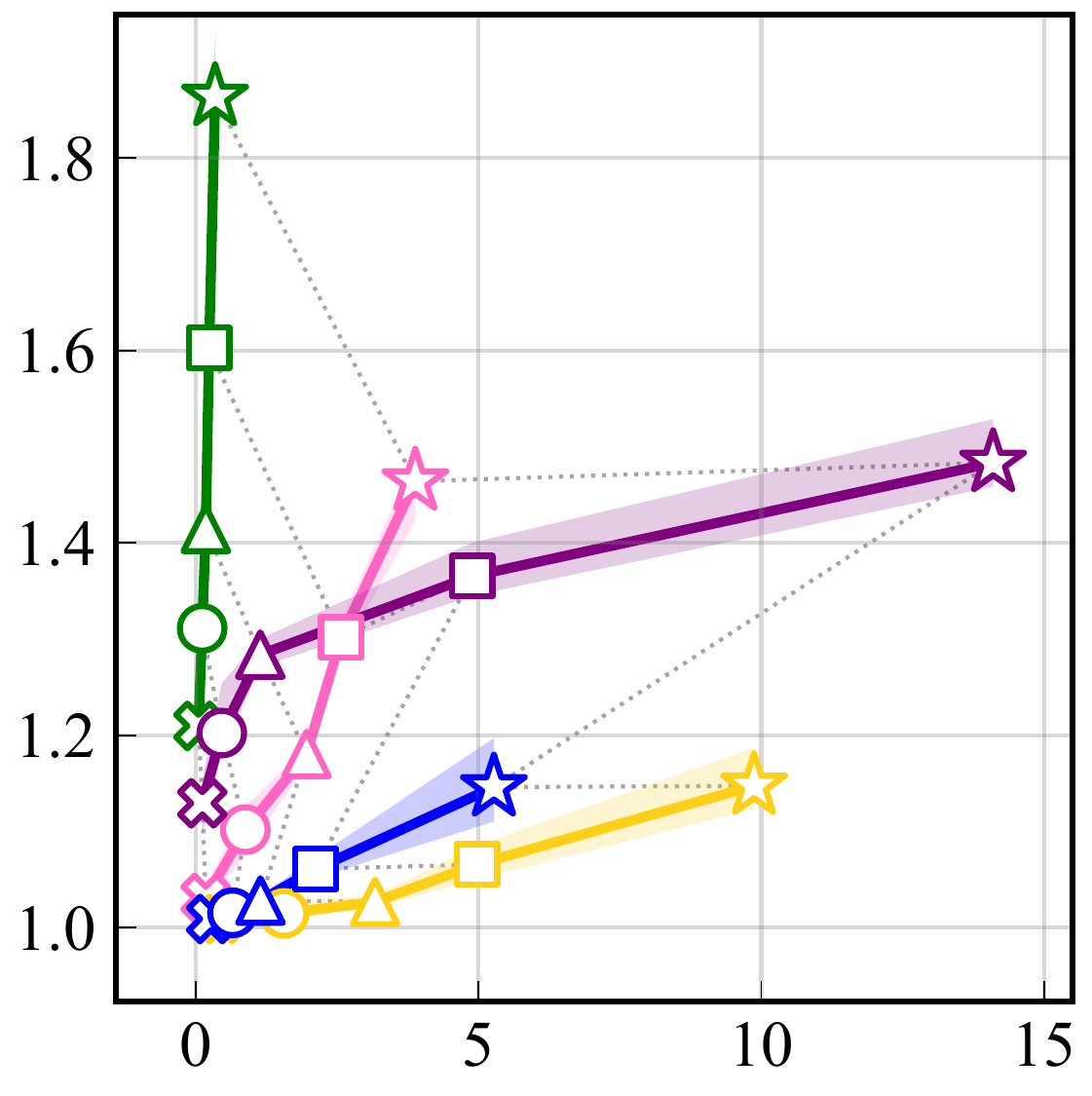}{194x194}{13,214}{}&
    \\
    \ylabel &
    \entry{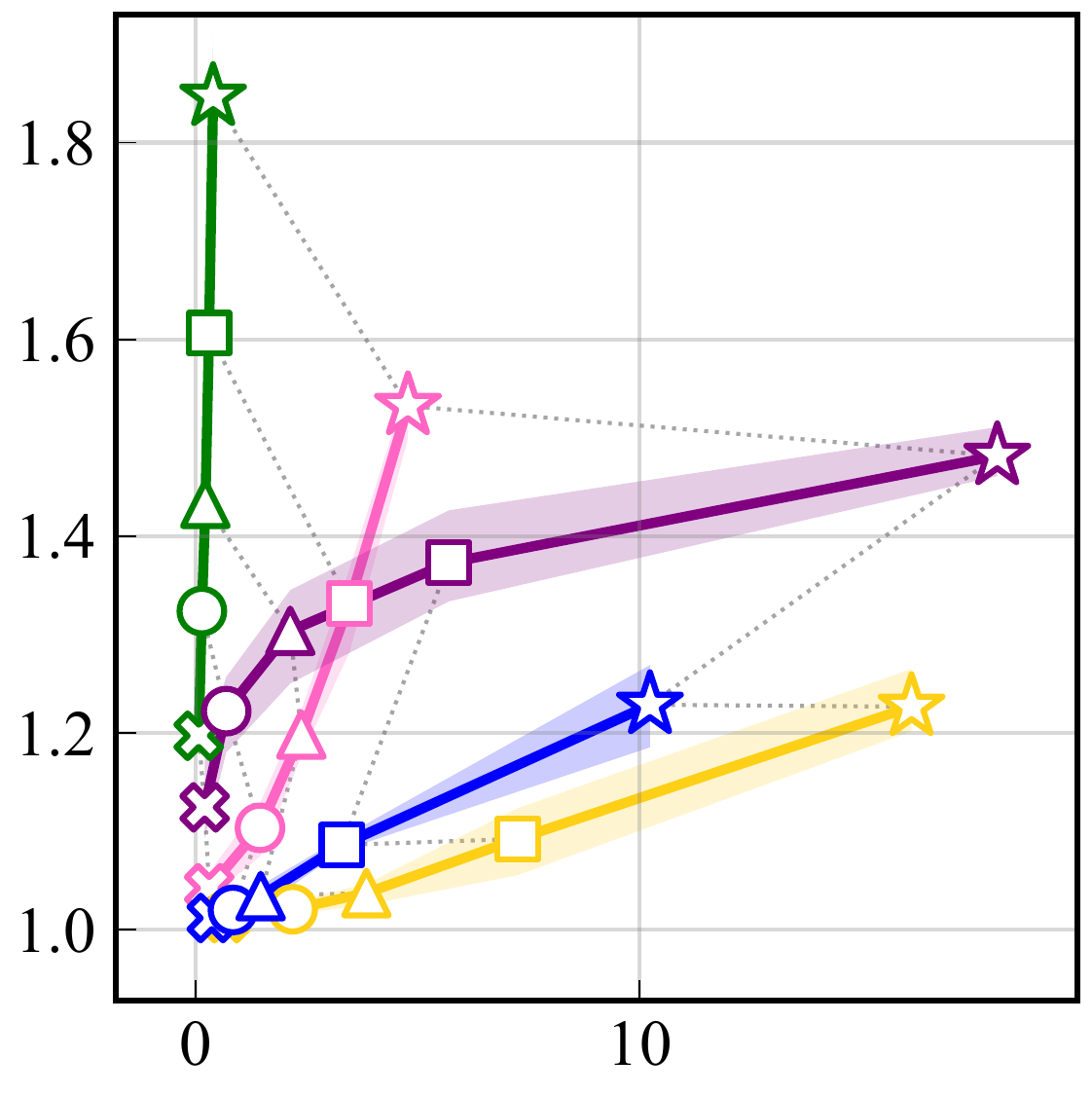}{194x194}{14,784}{}&
    \entry{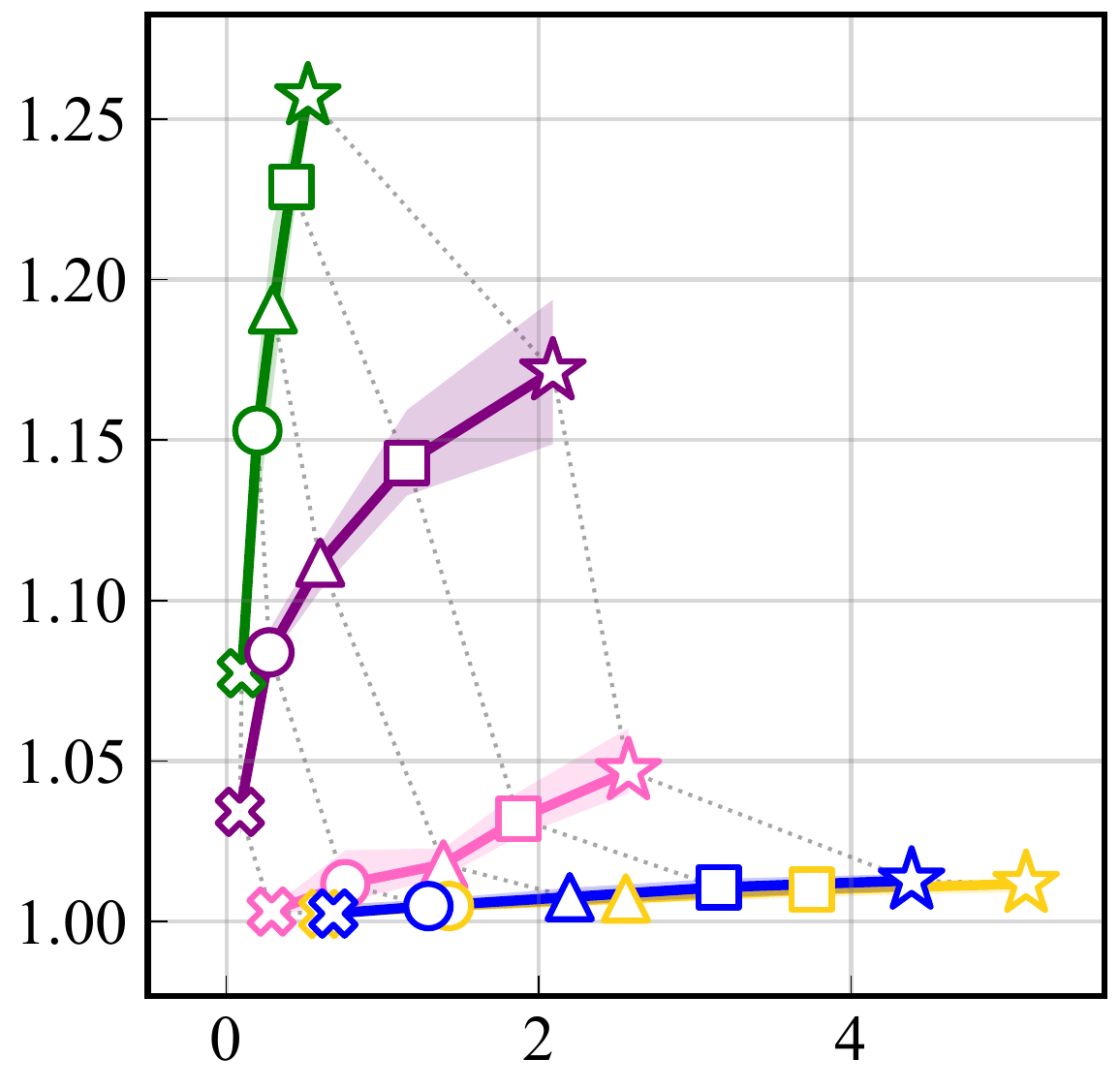}{256x256}{47,768}{}&
    \entry{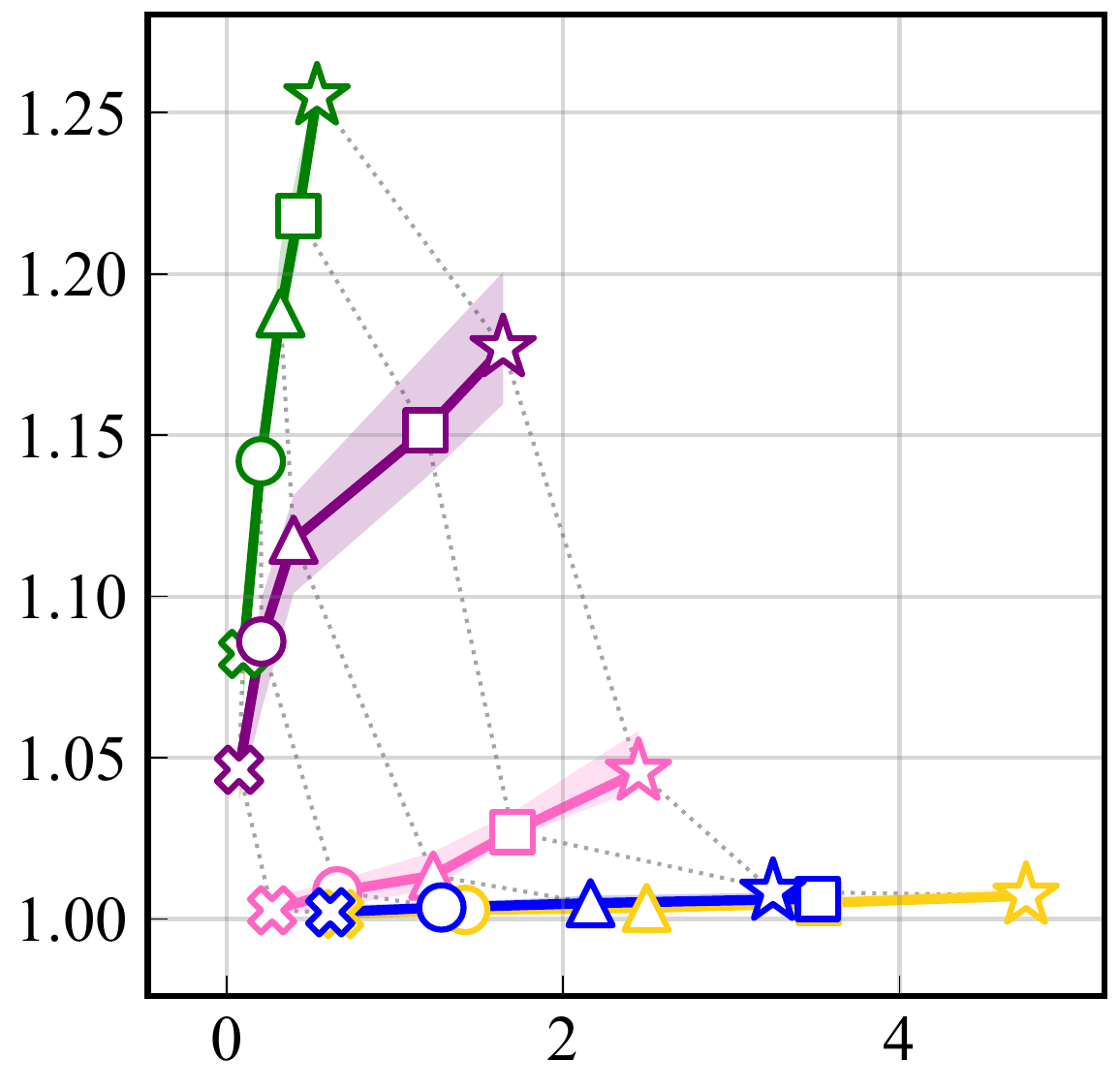}{256x256}{47,768}{}&
    \entry{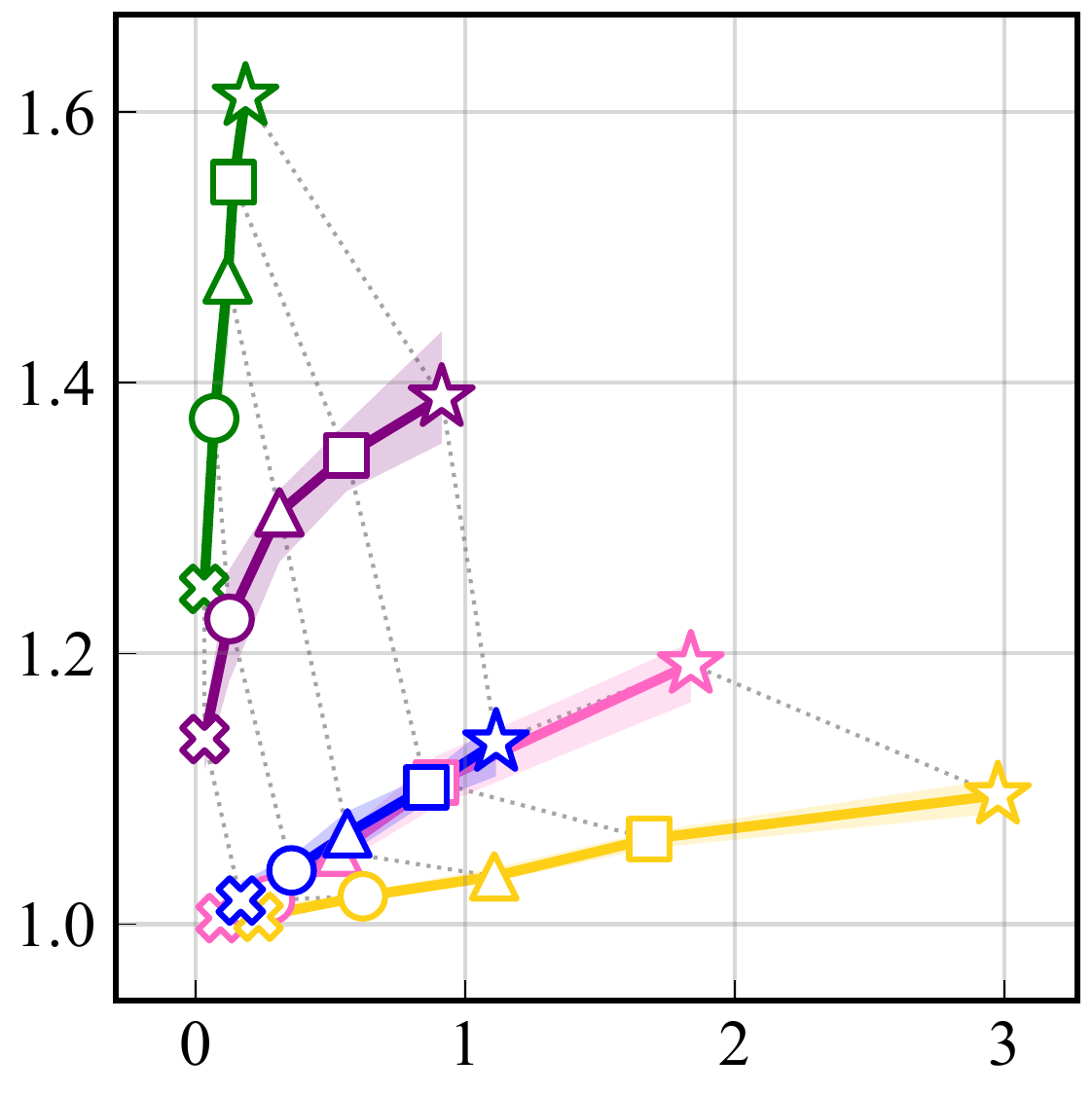}{170x84}{9,776}{}&
    \entry{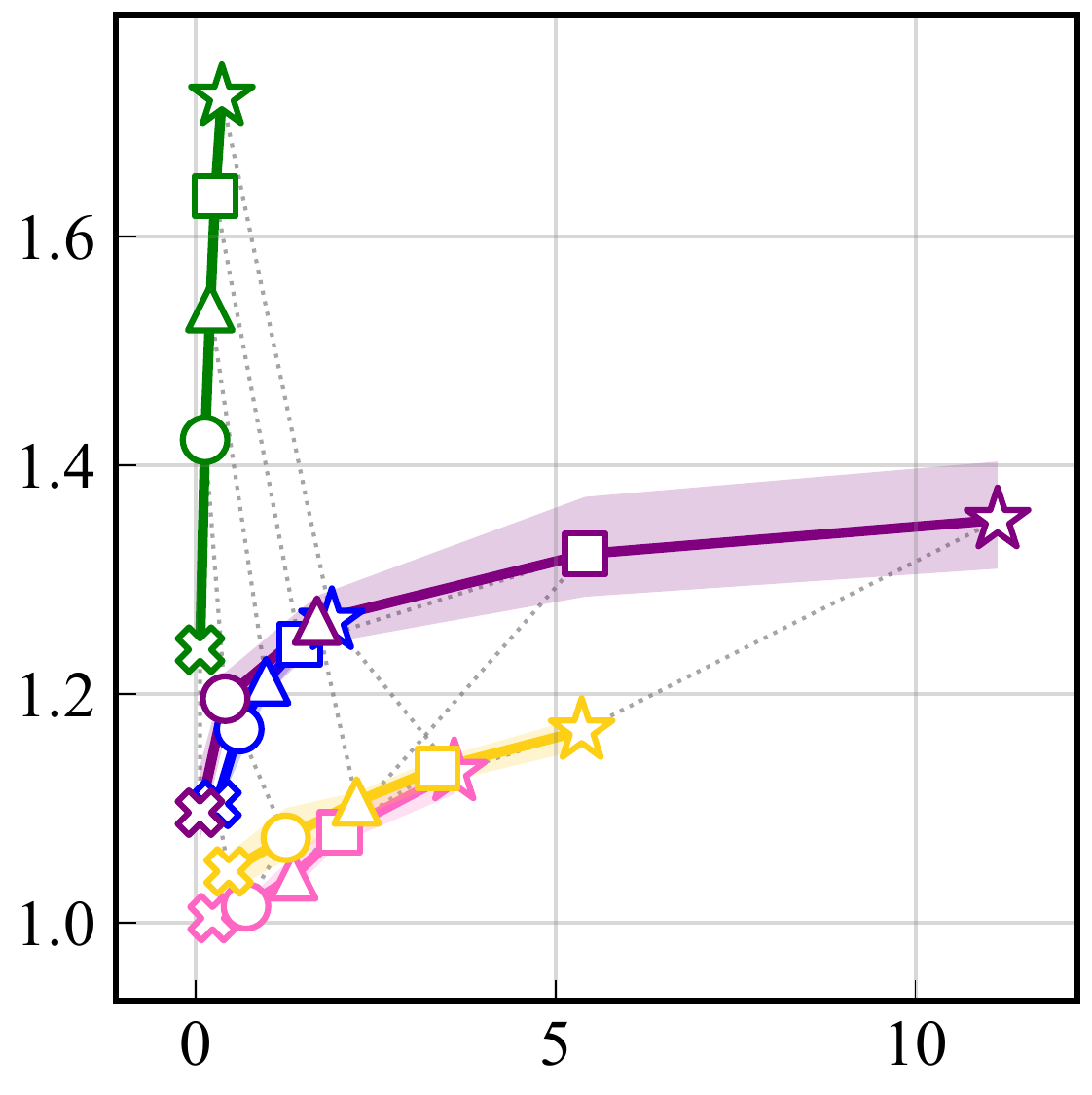}{321x123}{22,599}{1}&
    \\
    &\multicolumn{5}{c}{{\small $\leftarrow$ runtime [\SI{}{\second}]}}
    \\
    &\multicolumn{5}{c}{\includegraphics[width=0.5\linewidth,clip,trim={0 1.0cm 0 0.5cm}]{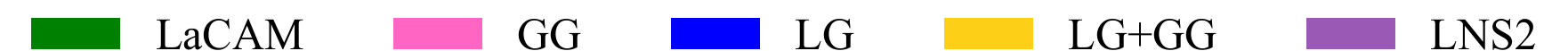}}
  \end{tabular}
  \caption{
  Performance evaluation of suboptimal MAPF methods, assessed by flowtime and runtime.
  \emph{Top:}
    Instance-wise comparison.
    From the MAPF benchmark, we extracted a total of 644 instances across five agent scales $\{200, 400, 600, 800, 1000\}$, selecting five instances per setting from 32 different grids.
    ``cost reduction'' is calculated using the LaCAM scores as a reference.
    The instances are sorted by the LG scores.
  \emph{Bottom:}
  Map-wise analysis with varying numbers of agents from $200$ to $1000$.
  Each subfigure reports the average over the solved instances among 25 cases per setting.
  Note that LaCAM-based solvers successfully solved all instances within the \SI{30}{\second} time limit, whereas LNS2 failed on \mapname{den312d} when $|A| \in \{800, 1000\}$.
  Flowtime scores are normalized by a trivial lower bound $\sum_{i\in A}\text{dist}(s_i, g_i)$.
  The number of vertices for each grid is shown with parentheses.
  }
  \label{fig:result-main}
\end{figure*}
}

\subsection{Qualitative Comparison}
We first show the effect of guidance visually, using a highly constrained environment \emph{maze-128-128-10}, in which agents can easily become congested, making it challenging to find near-optimal solutions.
\Cref{fig:heatmap} contrasts four LaCAM's solutions: one without guidance, with global guidance (GG), local guidance (LG), or both (LG+GG).

GG can provide coarse, time-independent information about congestion mitigation, resulting in diversified vertex usage in its solution, particularly in area-A.
This comes with a cost reduction from the original LaCAM.
Meanwhile, it fails to mitigate congestion in area-B, where most agents have a unique shortest start-goal path segment.

This is where LG stands out.
It quantitatively and qualitatively smooths out agent flow by providing more detailed, spatiotemporal cues to mitigate local congestion in bottleneck regions (e.g., area-B).
Indeed, LG discourages the planner from overusing specific vertices, resulting in a 38\% reduction in cost.
LG+GG inherits features from both LG and GG, resulting in slightly better performance than LG.
More examples appear in the Appendix.

{
\begin{figure}[t!]
\centering
\begin{tikzpicture}
\node[anchor=south] at (0, 0)
{\includegraphics[width=0.45\linewidth]{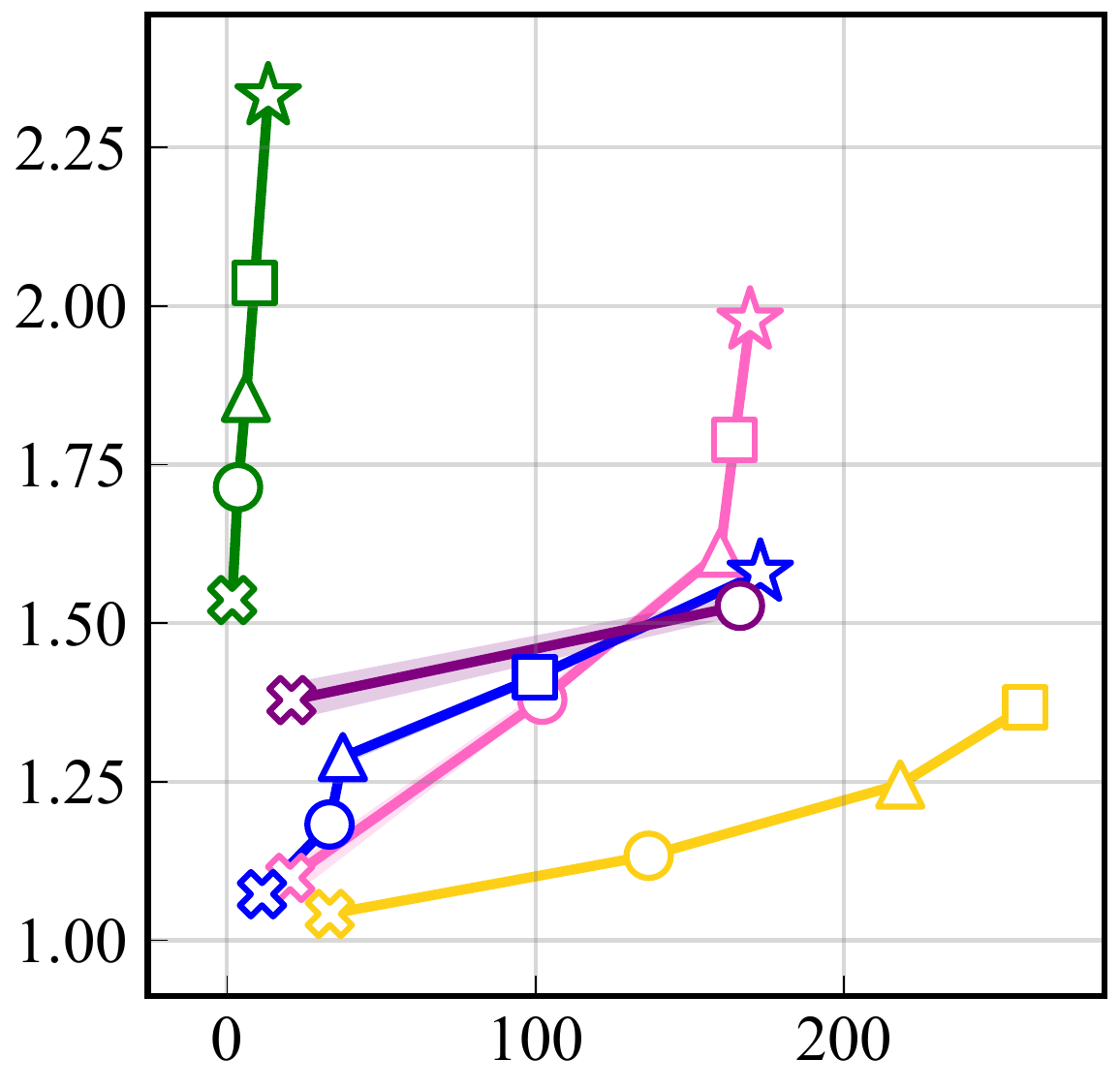}};
\node[] at (0, 0) {\small $\leftarrow$ runtime [\SI{}{\second}]};
\node[rotate=90] at (-0.27\linewidth, 0.25\linewidth) {\small $\leftarrow$ flowtime / LB};
\node[anchor=west] at (-0.9, 3.5) {\small \textcolor[HTML]{008000}{LaCAM}};
\node[anchor=west] at (0.2, 3.1)  {\small \textcolor[HTML]{FF66C4}{GG}};
\node[anchor=west] at (-0.7, 2.0)  {\small \textcolor[HTML]{800080}{LNS2}};
\node[anchor=west] at (0.8, 1.9)  {\small \textcolor[HTML]{0000FF}{LG}};
\node[anchor=west] at (0.5, 0.8)  {\small \textcolor[HTML]{F6D24B}{LG+GG}};
\begin{scope}[xshift=0.27\linewidth,yshift=0.15\linewidth]
  \draw[] (0, 0) --
  node[pos=0]{\small $\times$}
  node[pos=0.25]{$\bullet$}
  node[pos=0.5]{$\blacktriangle$}
  node[pos=0.75]{\scriptsize $\blacksquare$}
  node[pos=1.0]{\scriptsize $\bigstar$}
  node[right,pos=0]{\footnotesize 2,000}
  node[right,pos=0.25]{\footnotesize 4,000}
  node[right,pos=0.5]{\footnotesize 6,000}
  node[right,pos=0.75]{\footnotesize 8,000}
  node[right,pos=1]{\footnotesize 10,000}
  node[above right=0.5,pos=1.05]{\footnotesize agents}
  (0.2, 1.8);
\end{scope}
\end{tikzpicture}
\caption{
  Scalability test on \mapname{warehouse-20-40-10-2-2}.
}
\label{fig:scalability}
\end{figure}
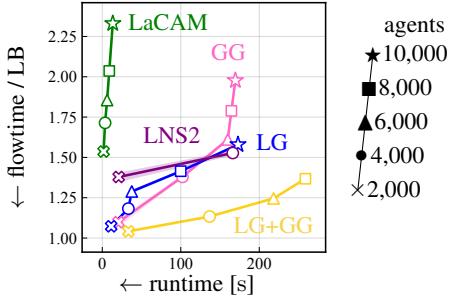
}

{
\setlength{\tabcolsep}{2pt}
\newcommand{\entry}[5]{
\begin{minipage}{0.47\linewidth}
\centering
\scriptsize
\begin{tikzpicture}
\node[anchor=north] at (0, 0)
{\includegraphics[width=1.0\linewidth,height=1\linewidth,clip,trim={0.0cm 0.0cm 0.0cm 0.0cm}]
{fig/raw/collision_cost/collision_cost_#1_#5}};
\node[anchor=south] at (0.3, -0.1) {\mapname{#5}, $|A|$$=$#2};
\end{tikzpicture}
\end{minipage}
}
\begin{figure}[th!]
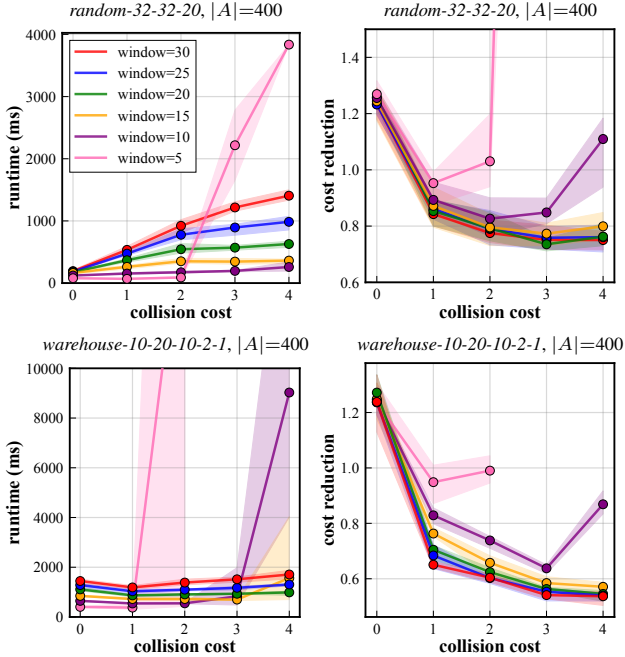

  \centering
  \begin{tabular}{cc}
    \entry{runtime}{400}{0.8,-2.2}{0.3,-3}{random-32-32-20} &
    \entry{cost}{400}{0.8,-2.4}{0.3,-3.3}{random-32-32-20} \\
    \entry{runtime}{400}{0.8,-2.2}{0.3,-3}{warehouse-10-20-10-2-1} &
    \entry{cost}{400}{0.8,-2.4}{0.3,-3.3}{warehouse-10-20-10-2-1} \\
  \end{tabular}
\caption{
Effect of the collision cost $\alpha$ and window size $w$.
``cost reduction'' represents the ratio from LaCAM.
}
\label{fig:collision_cost}
\end{figure}
}

\subsection{Quantitative Comparison}
We next evaluate the guidance effect quantitatively through various maps.
As summarized in \cref{fig:result-main}, overall, the proposed LG substantially reduces the solution cost.
With a few exceptions, this reduction exceeds that of GG.
Although LG incurs higher computational overhead than GG---due to the guidance reconstruction at every configuration generation---the additional runtime remains within a few seconds in most cases, even with $1,000$ agents.
We argue that this overhead is a worthwhile trade-off for the marked improvement in solution quality.
Notably, LG outperforms LNS2 in most scenarios, delivering better solutions in less time.
These results suggest that local guidance establishes a new Pareto frontier in MAPF, balancing speed and solution quality.

Broadly, LG achieves compelling performance, while GG surpasses LG in \mapname{warehouse-20-40-10-2-1}.
This is because GG effectively streamlines agent traffic flow, particularly in workspaces with many narrow corridors, by suppressing wasteful back-and-forth movements within them (see also the Appendix, showing the vertex-usage heatmap like~\cref{fig:heatmap}).
In contrast, results from the other maps indicate that mitigating local congestion is more effective in general.
Combining LG and GG compensates for each other's shortcomings and yields the best solution quality among all tested solvers.
However, this comes at the cost of additional computation for both global and local guidance, and the performance gain over LG alone remains modest.
Developing more effective integration of global and local guidance remains an important direction for future work.

\paragraph{Scalability.}
\Cref{fig:scalability} evaluates the scalability of LG with up to $10{,}000$ agents on the warehouse map, under an extended time limit of \SI{300}{\second}.
While the fast construction of GG at this scale has been reported as challenging~\cite{okumura2024lacam3}, LG effectively reduces the solution cost (by approximately $30\%$) with moderate planning time, remaining significantly faster than both GG and LNS2.
These results suggest that the notion of local guidance is effective not only for practical problem sizes, as shown in \cref{fig:result-main}, but also remains effective in ultra-large-scale MAPF scenarios.

\subsection{Design Justification}

\paragraph{Collision Cost and Window Size.}
The coefficient $\alpha$ used in \cref{eq:cost} to penalize collisions affects both solution quality and the runtime, together with window size $w$, as illustrated in \cref{fig:collision_cost}, which shows two typical cases.
There is a sweet spot;
overweighting collisions may worsen the resulting solutions because agents are encouraged to be conservative to avoid them, while underweighting yields faster planning but leaves room for improvement in congestion mitigation.
Likewise, enlarging the window size can improve the quality of the solution up to a certain extent, but at the expense of computation time.

{
\setlength{\tabcolsep}{2pt}
\newcommand{\entry}[4]{
\begin{minipage}{0.47\linewidth}
\centering
\scriptsize
\begin{tikzpicture}
\node[anchor=north] at (0, 0)
{\includegraphics[width=1.0\linewidth,height=1\linewidth]{fig/raw/refinement/refine_#1}};
\node[anchor=south] at (0.3, -0.1) {\mapname{random-32-32-20}};
\end{tikzpicture}
\end{minipage}
}

\begin{figure}[t!]
  \centering
  \begin{tabular}{cccccc}
    \entry{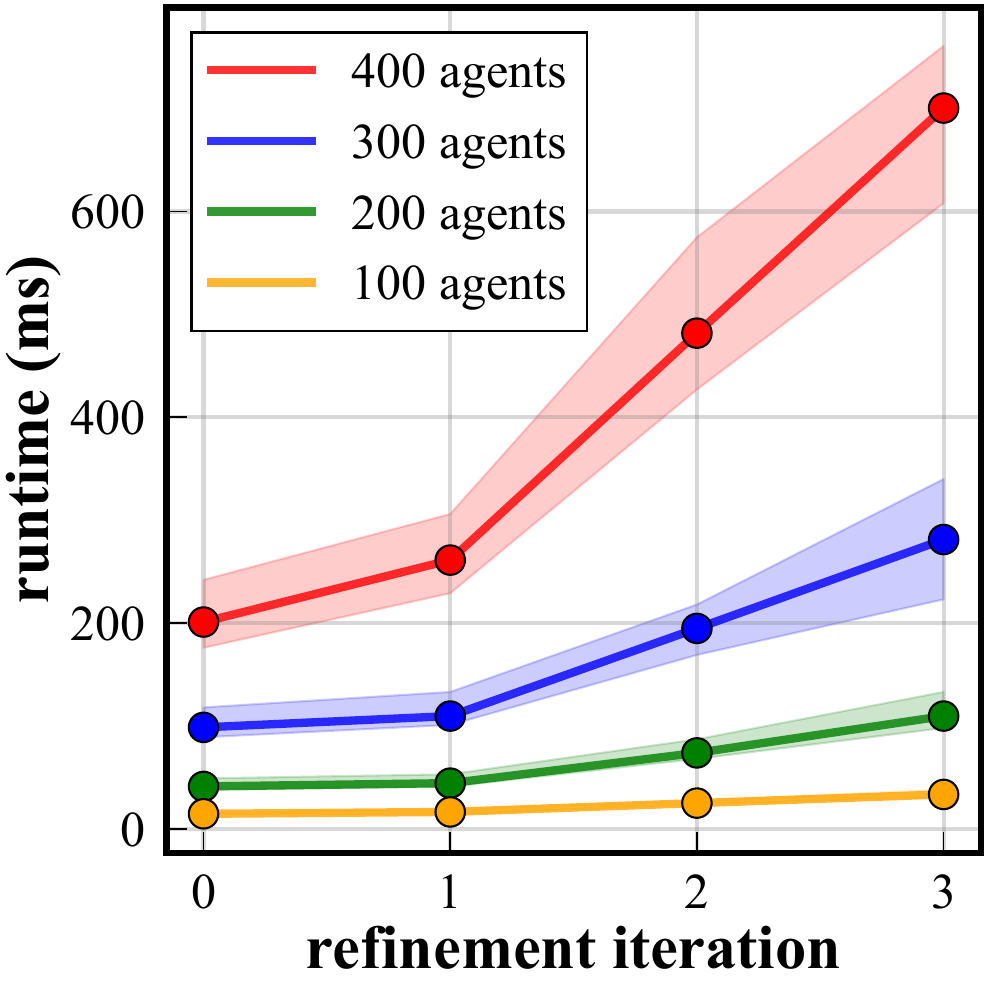}{400}{0.8,-2.2}{0.3,-3}&
    \entry{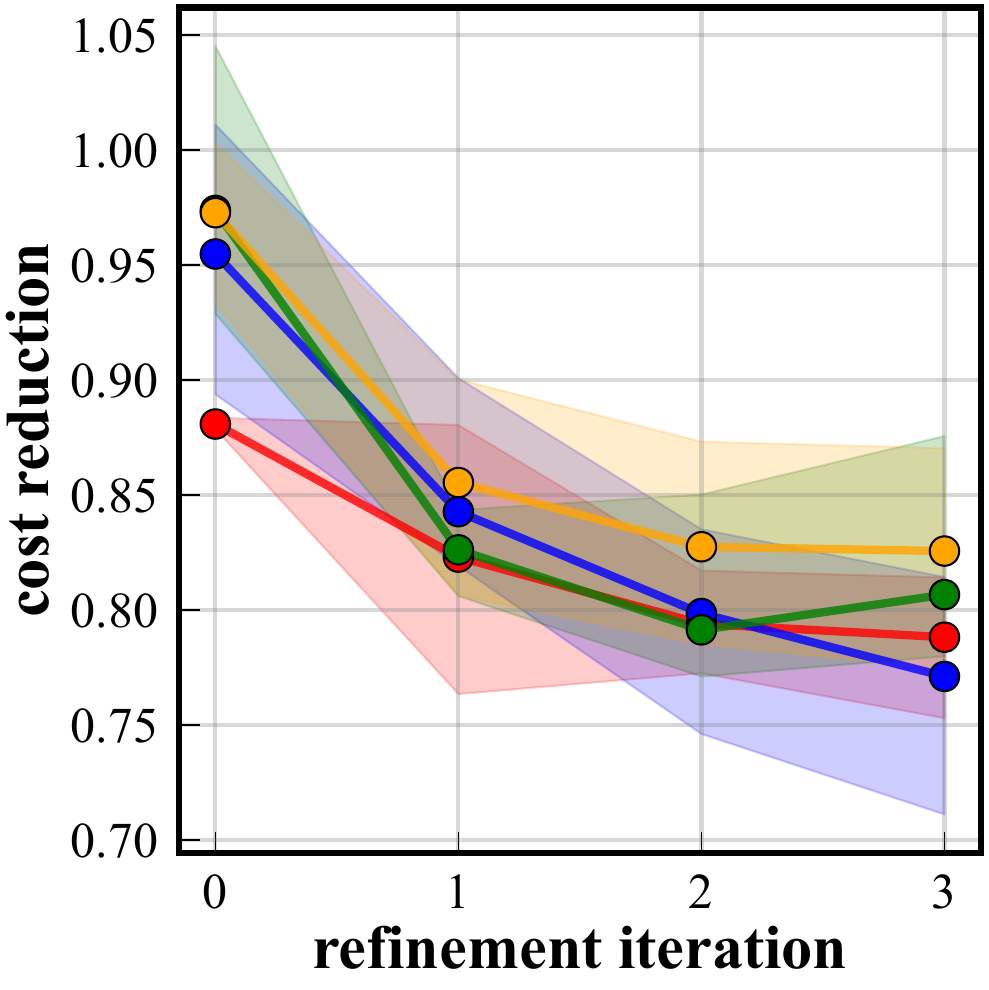}{400}{0.8,-2.4}{0.3,-3.3}\\
  \end{tabular}
\caption{
Effect of the plan iteration and initialization.
``0'' plans the paths from scratch without further refinement.
``1'' uses the path cache and refines it once, and so on.
}
\label{fig:refinement}
\end{figure}
}

{
\setlength{\tabcolsep}{0pt}
\newcommand{\entry}[4]{
\begin{minipage}{0.31\linewidth}
\centering
\scriptsize
\begin{tikzpicture}
\node[anchor=north] at (0, 0)
{\includegraphics[width=0.9\linewidth,height=1.5\linewidth,clip,trim={1.2cm 1.2cm 0.0cm 0.0cm}]{fig/raw/sort/sort_#1}};
\node[anchor=south, align=center] at (0.3, -0.1) {%
  \mapname{#1}\\
  $|A|$$=$#2
};
\node[anchor=south west] at (#3) {\textcolor{red}{unsorted}};
\node[anchor=south east] at (#4) {\textcolor{blue}{sorted}};
\end{tikzpicture}
\end{minipage}
}
\begin{figure}[t!]
  \centering
  \begin{tikzpicture}
    \node (imgs) {
      \begin{tabular}{ccc}
        \entry{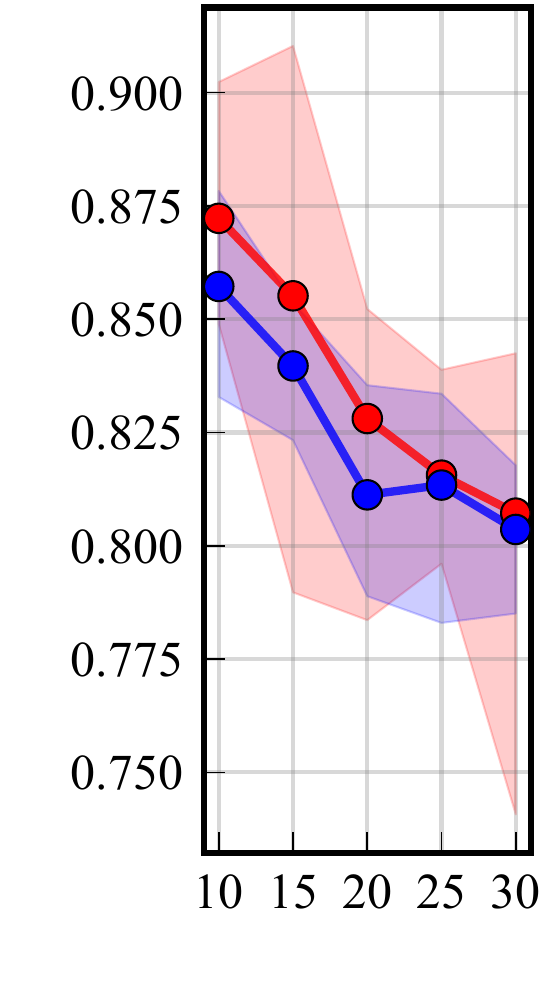}{400}{0.1,-1.6}{0.3,-3.0} &
        \entry{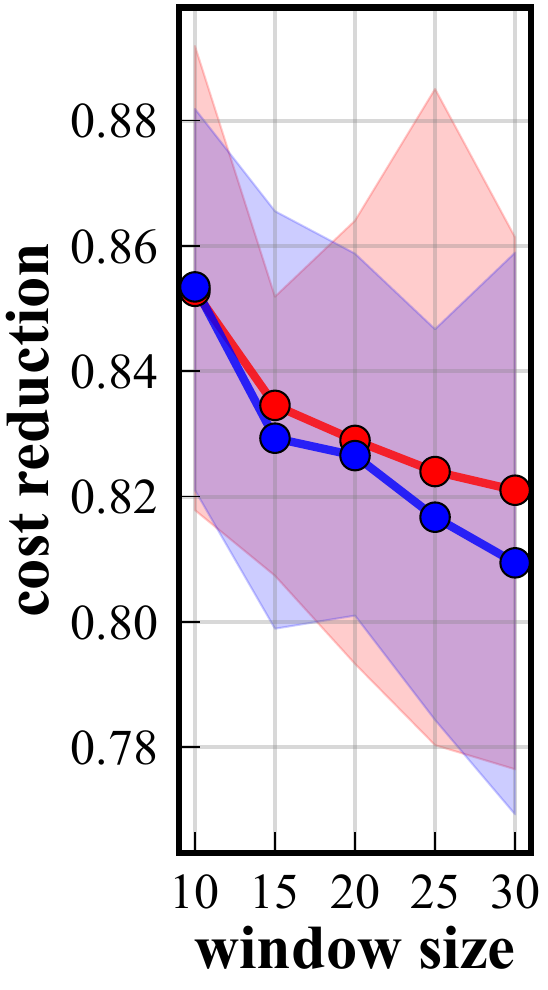}{400}{-0.1,-1.6}{0.3,-3.0} &
        \entry{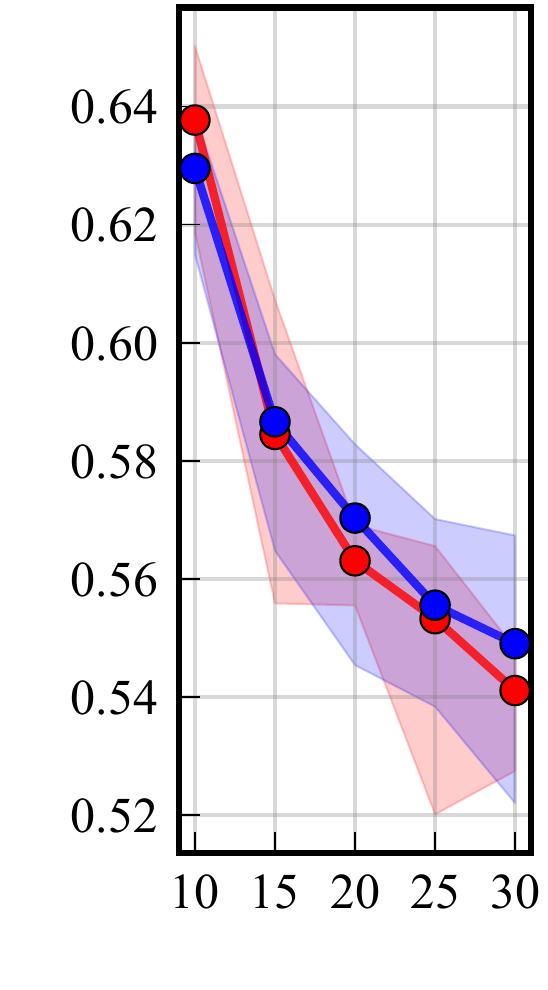}{400}{-0.6,-3.5}{1.0,-2.0} \\
      \end{tabular}
    };
    \node[anchor=south] at (imgs.south) [yshift=-0.9em] {\small \textbf{window size}};
    \node[anchor=east, rotate=90] at (imgs.west) [xshift=2.0em] {\small \textbf{cost reduction}};
  \end{tikzpicture}
  \caption{
  Effect of the agent order used in the guidance construction.
  The runtime difference is negligible and omitted.
  }
  \label{fig:result-sort}
\end{figure}
}

{
\setlength{\tabcolsep}{2pt}
\newcommand{\entry}[4]{
\begin{minipage}{0.47\linewidth}
\centering
\scriptsize
\begin{tikzpicture}
\node[anchor=north] at (0, 0)
{\includegraphics[width=1.0\linewidth,height=1\linewidth]{fig/raw/k_step_update/k_step_update_#1_28}};
\node[anchor=south] at (0.3, -0.1) {\mapname{random-32-32-20}, $|A|$$=$#2};
\end{tikzpicture}
\end{minipage}
}
\begin{figure}[tp!]
  \centering
  \begin{tabular}{cccccc}
    \entry{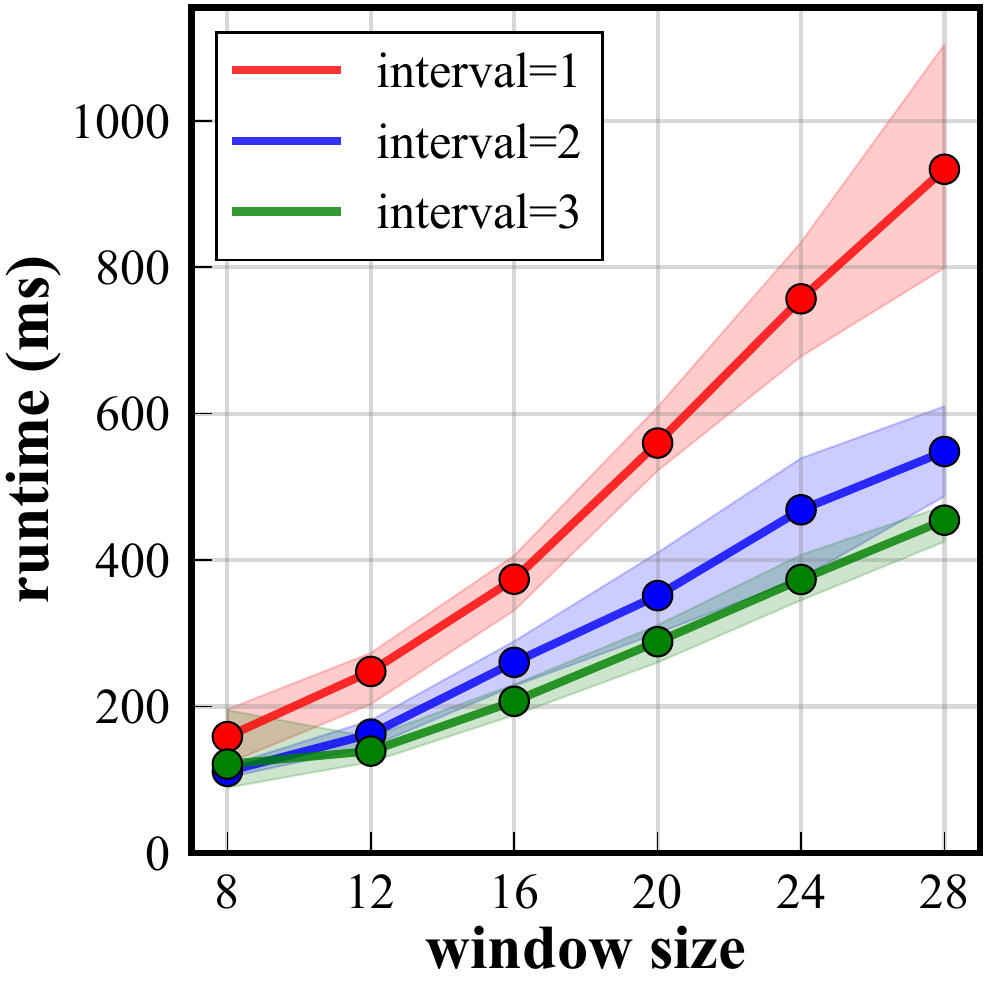}{400}{0.8,-2.2}{0.3,-3}&
    \entry{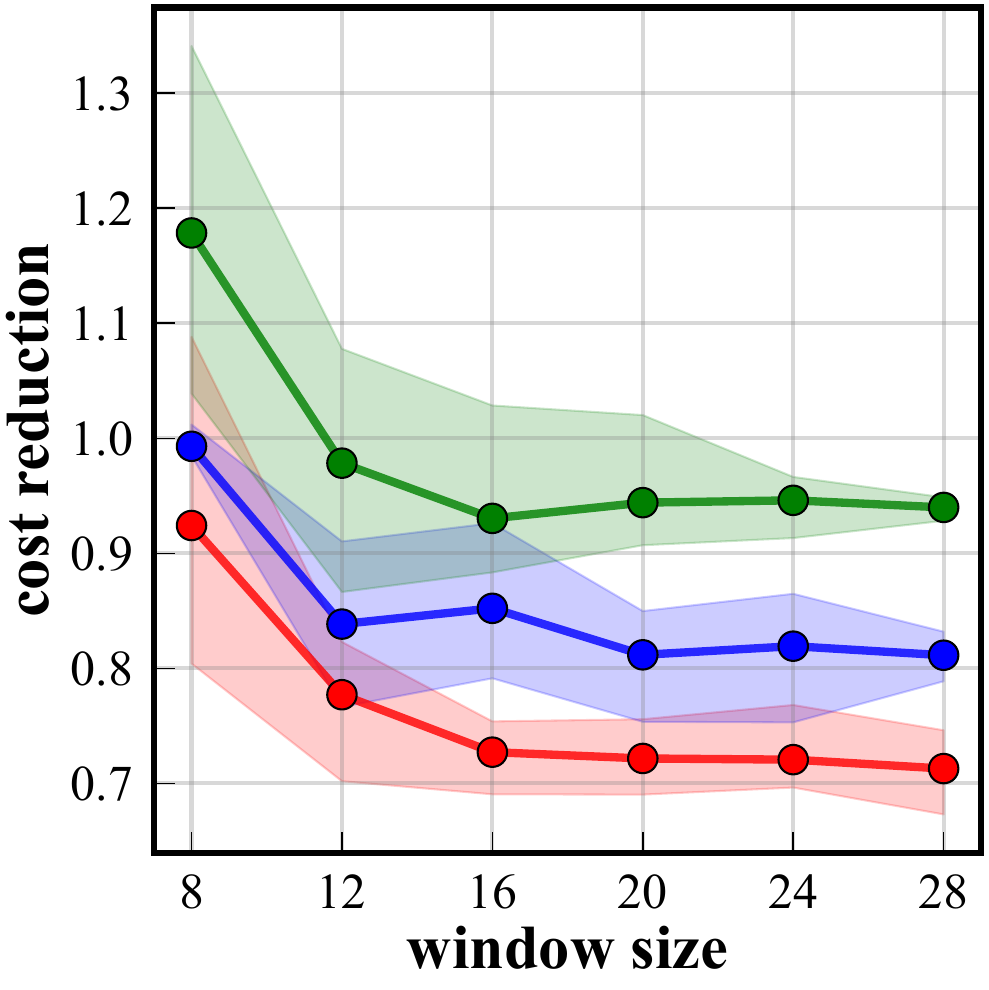}{400}{0.8,-2.4}{0.3,-3.3}\\
  \end{tabular}
\caption{
Effect of the frequency of the local guidance update.
``1'' updates every configuration generation, ``2'' updates every two generations, and so on.
}
\label{fig:k_step_update}
\end{figure}
}

\paragraph{Plan Iteration and Initialization.}
The guidance construction process is essentially non-stationary, as the guidance path for each agent changes in consideration of the others.
We tackle this issue by caching the previous result and using it for bootstrapping.
\Cref{fig:refinement} demonstrates its effect.
Our results so far are obtained with one refinement iteration, while removing the cache, which corresponds to zero refinement, significantly drops the performance.
Further refining the guidance can improve its quality, but the benefits gradually diminish, and the computation time becomes costly.

\paragraph{Agent Order.}
When constructing the guidance, the agent order  (\cref{algo:guidance:order} in \cref{algo:guidance}) has a design choice.
We sort the agents according to the number of collisions as described.
\Cref{fig:result-sort} illustrates the typical behavior, showing performance improvements across varying window sizes, compared to the unsorted case.
Unlike PP, whose performance greatly depends on the ordering, this modest effect occurs as the guidance \emph{indirectly} facilitates collision-free path computation.
Meanwhile, the \mapname{warehouse-10-20-10-2-1} map is an exception;
it contains numerous narrow passages, which may cause the agents to exhibit oscillatory behavior with the sorting operation.
We note that the differences in computation time are negligible.

{
\setlength{\tabcolsep}{1pt}
\newcommand{\entry}[2]{
\begin{minipage}{0.19\linewidth}
  \centering
  \begin{tikzpicture}
    \node[anchor=north east] at (0, 0)
    {\includegraphics[width=1\linewidth,height=0.5\linewidth]{fig/raw/lns/lns_sub_#1_#2_agents}};
    \node[anchor=north east] at (-0.1, -0.1) {\scriptsize #2 agents};
  \end{tikzpicture}
\end{minipage}
}
\renewcommand*{\arraystretch}{0.8}
\newcommand{\ylabel}{\rotatebox{90}{\scriptsize \hspace{-2.3em}{flowtime / LB}}}
\begin{figure*}[tp!]
\centering
\begin{tabular}{cccccc}
  &\multicolumn{5}{c}{\scriptsize \mapname{random-64-64-20}}\vspace{-0.15cm}
  \\
  \ylabel&
  \entry{random_64_64_20}{200}&
  \entry{random_64_64_20}{400}&
  \entry{random_64_64_20}{600}&
  \entry{random_64_64_20}{800}&
  \entry{random_64_64_20}{1000}
  \\
  &\multicolumn{5}{c}{\scriptsize \mapname{warehouse_20_40_10_2_1}}\vspace{-0.15cm}
  \\
  \ylabel&
  \entry{warehouse_20_40_10_2_1}{200}&
  \entry{warehouse_20_40_10_2_1}{400}&
  \entry{warehouse_20_40_10_2_1}{600}&
  \entry{warehouse_20_40_10_2_1}{800}&
  \entry{warehouse_20_40_10_2_1}{1000}
  \\
  &\multicolumn{5}{c}{\scriptsize{runtime [\SI{}{\second}]}}
\end{tabular}
\includegraphics[width=0.45\linewidth,clip,trim={0.0cm 0.3cm 0.0cm 0.2cm}]{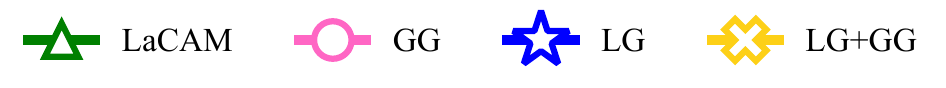}
\caption{
  Anytime MAPF performance by LaCAM (+guidance) followed by LNS.
}
\label{fig:lns}
\end{figure*}
}

{
\setlength{\tabcolsep}{1pt}
\newcommand{\entry}[4]{
\begin{minipage}{0.09\linewidth}
  \centering
  \begin{tikzpicture}
  \node[anchor=south west] at (0, 0)
  {\includegraphics[width=1.0\linewidth,height=2.0\linewidth]{fig/raw/lg_lacam_vs_lacam3/lacam_benchmark_#1}};
  \node[black,anchor=west,rotate=90] at (1.55, 0.22) {\tiny \mapname{#1}};
  \ifthenelse{\equal{#4}{1}}{
  \node[anchor=west] at (0.57, 0.15) {\tiny instances};
  }{}
  \end{tikzpicture}
\end{minipage}
}
\newcommand{\ylabel}{\rotatebox{90}{\scriptsize \hspace{-2.5em}{flowtime / LB}}}
\begin{figure*}[ht!]
  \centering
  \begin{tabular}{ccccccccccccc}
    \ylabel &
    \entry{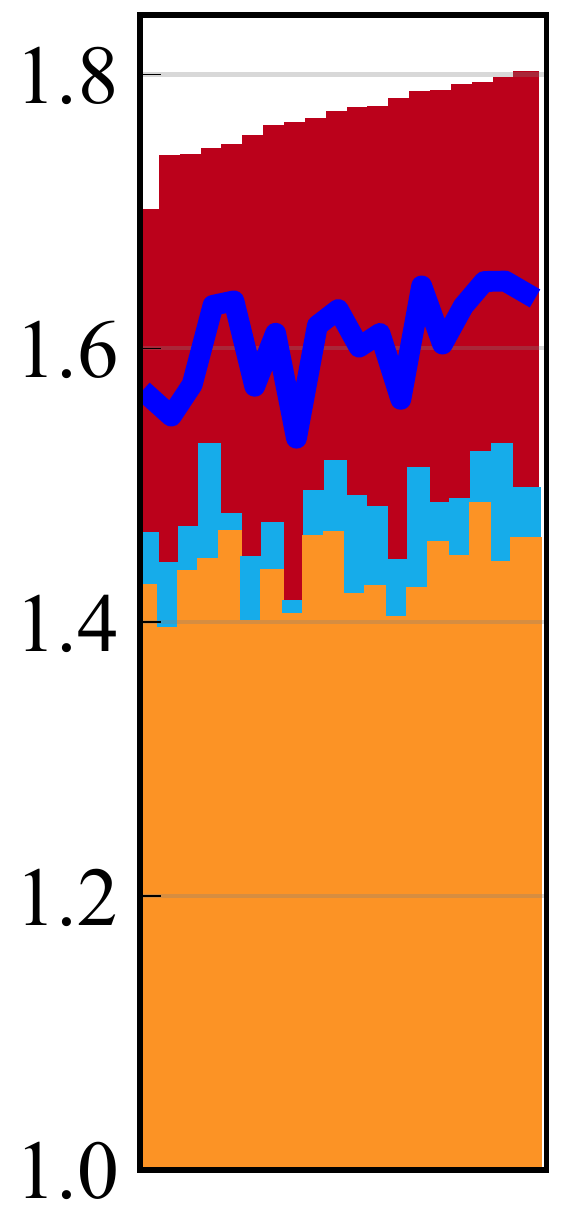}{48x48}{2,304}{1}&
    \entry{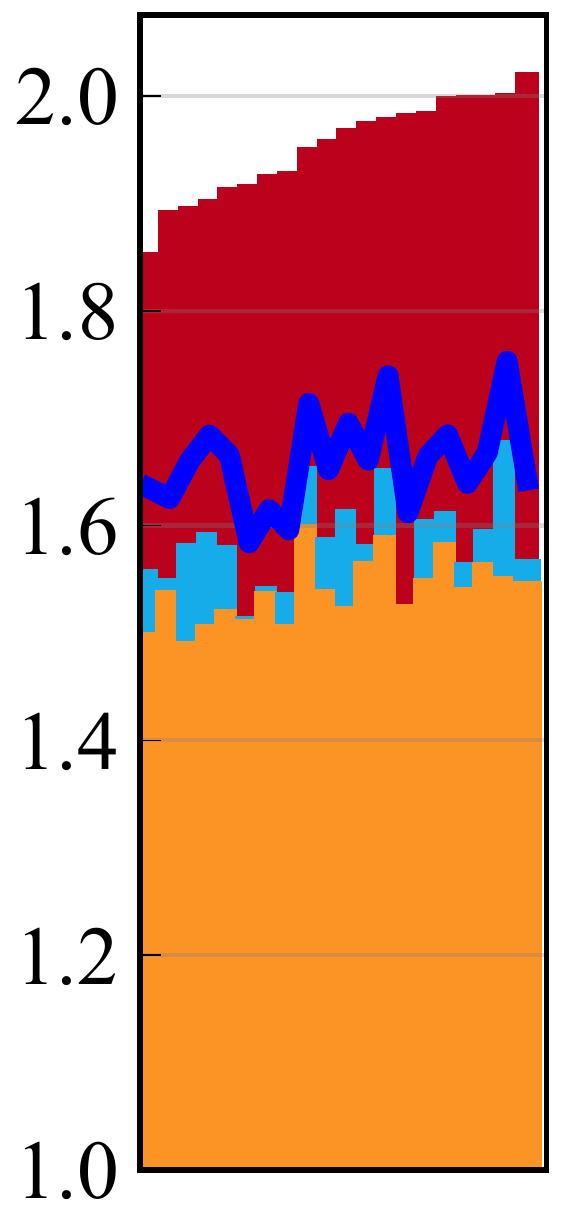}{64x64}{3,687}{}&
    \entry{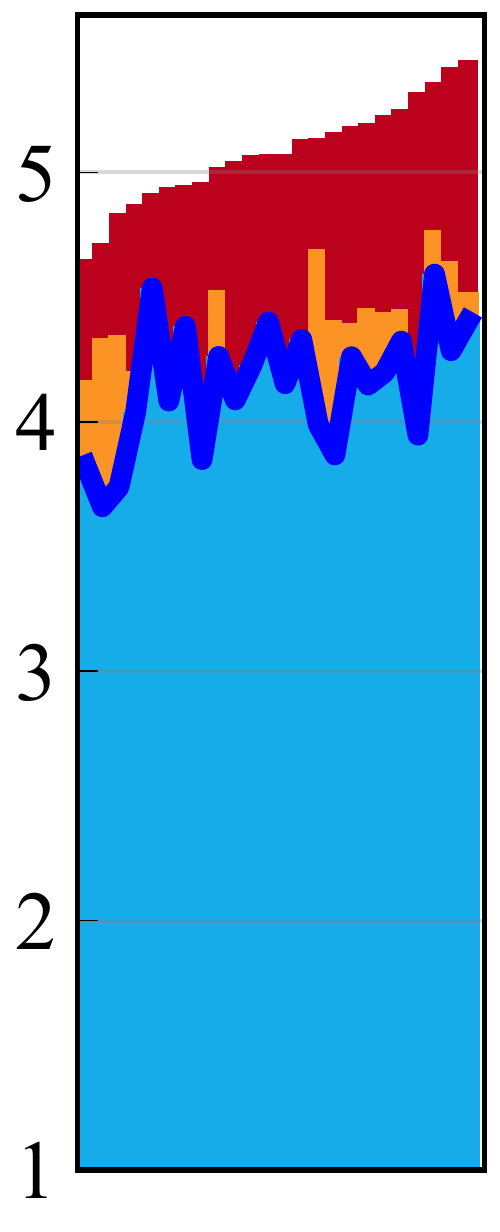}{65x81}{2,445}{}&
    \entry{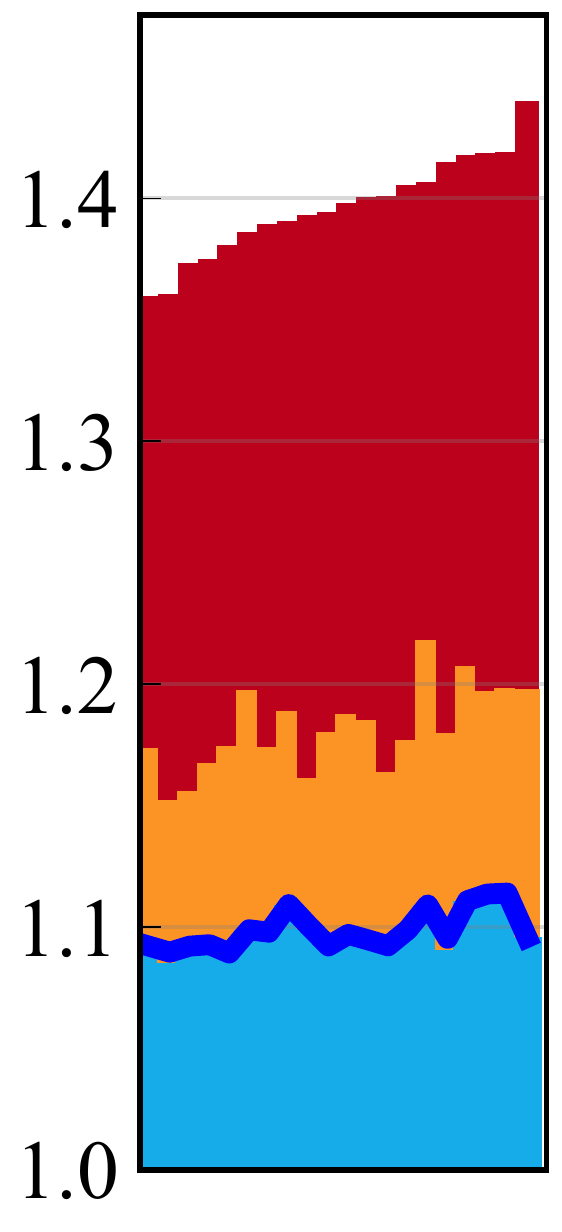}{128x128}{14,818}{}&
    \entry{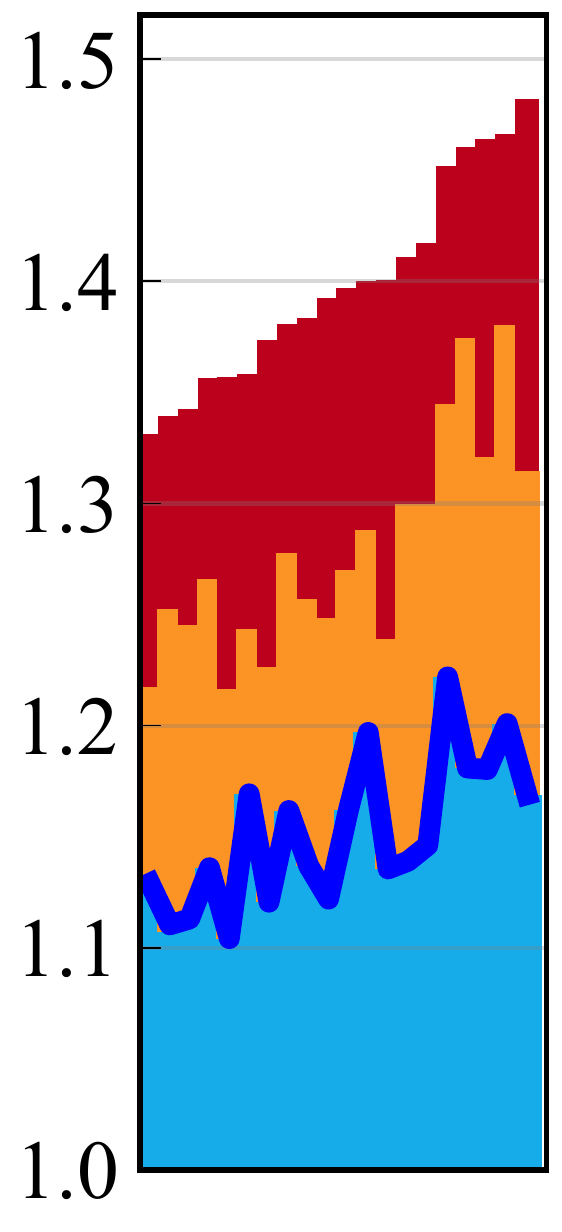}{194x194}{13,214}{}&
    \entry{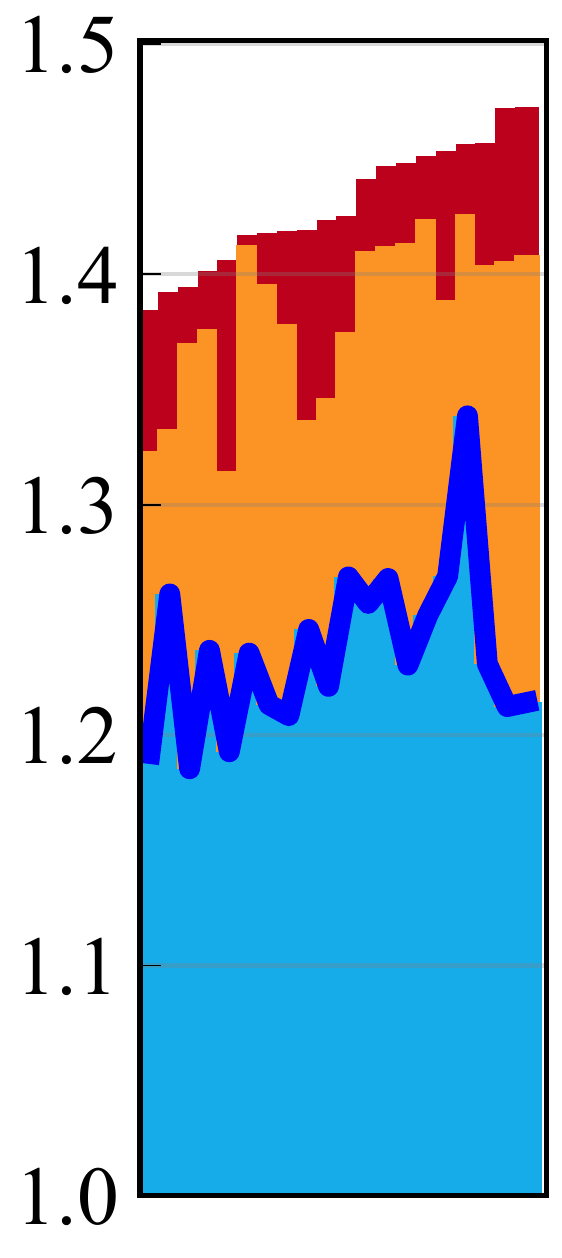}{194x194}{14,784}{}&
    \entry{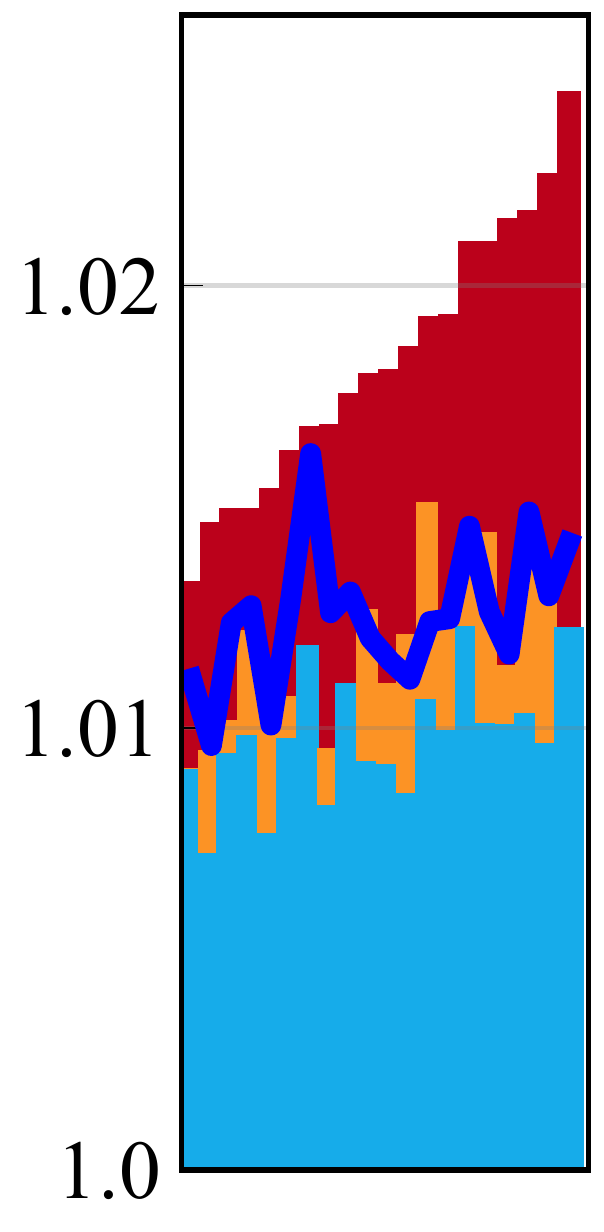}{256x256}{47,768}{}&
    \entry{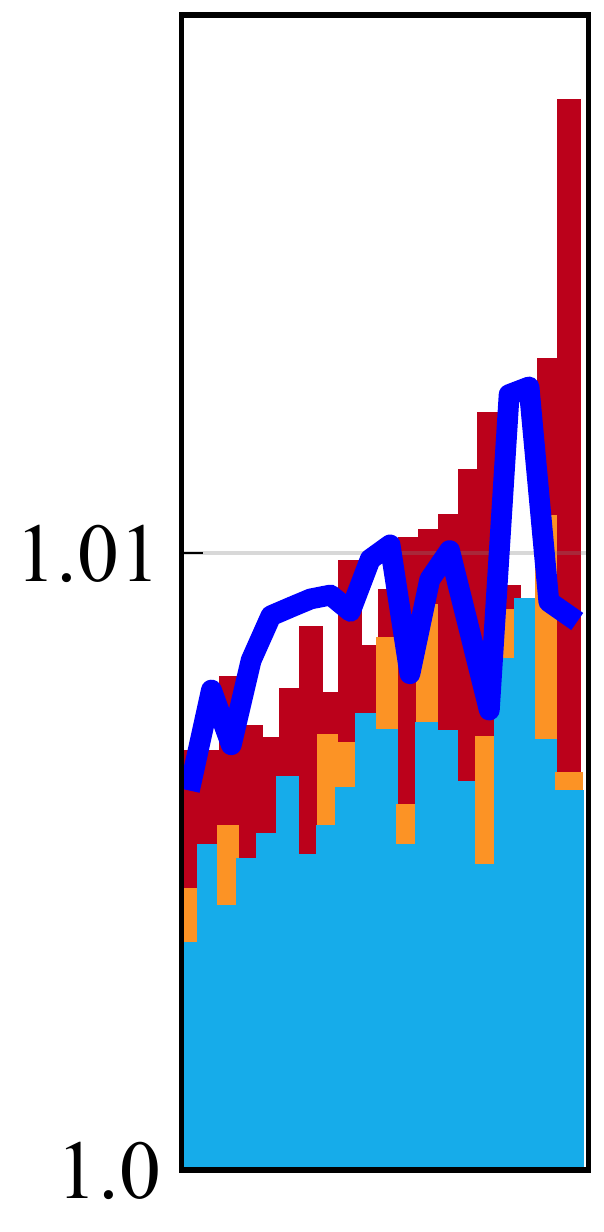}{256x256}{47,768}{}&
    \entry{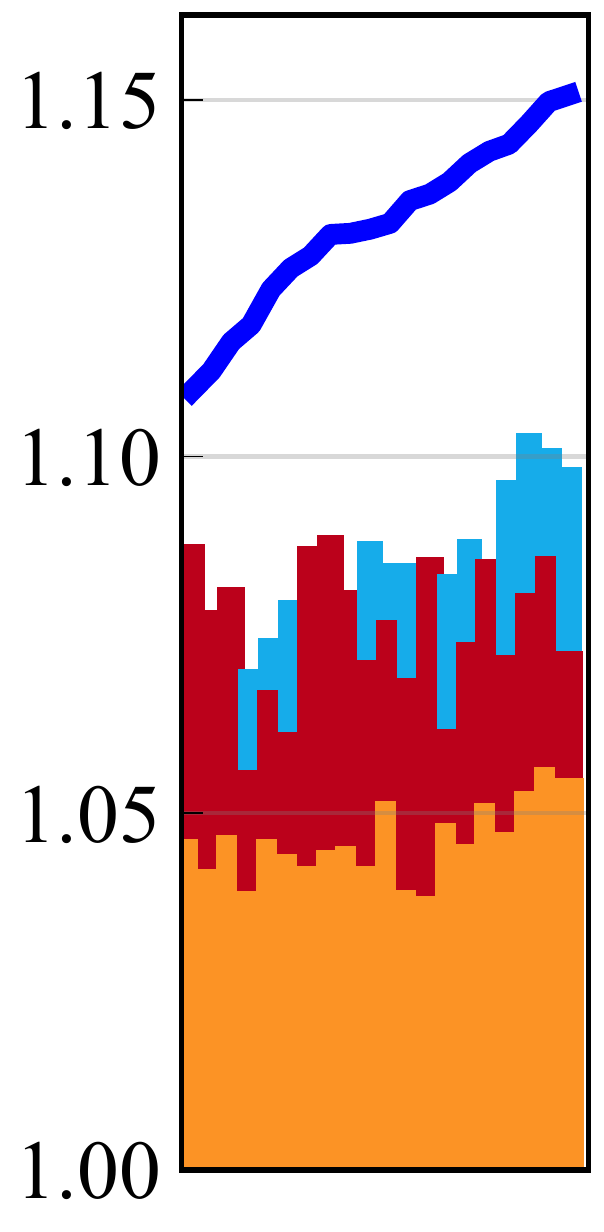}{170x84}{9,776}{}&
    \entry{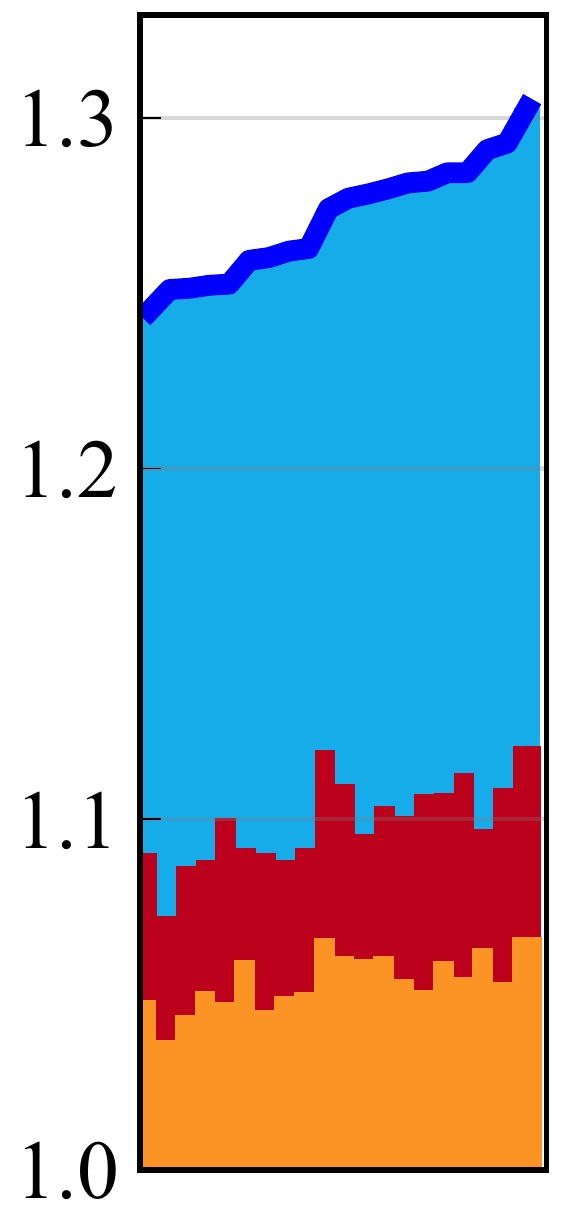}{321x123}{22,599}{}
    \\
    &\multicolumn{10}{c}{\includegraphics[width=0.6\linewidth,clip,trim={0 0.8cm 0 0.6cm}]
      {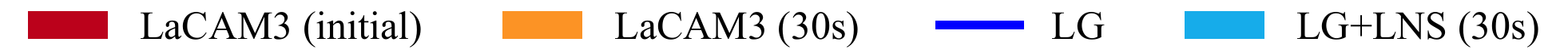}}
  \end{tabular}
\caption{
Instance-wise comparison with \texttt{lacam3}, using 25 instances with $1000$ agents for each map.
}
\label{fig:lacam3}
\end{figure*}
}

\paragraph{Less frequent guidance update is ineffective.}
One might think that local guidance construction is not required for each PIBT call.
Instead, it would be more computationally attractive if we could update the guidance less frequently.
Unfortunately, \cref{fig:k_step_update} shows that this is not the case.
This compares local guidance update frequencies: updating every configuration generation (default), every two generations, and every three.
Although there is a speed gain, the results show that less frequent updates could worsen performance compared to the original LaCAM.
Instead, updating the guidance every time with smaller window sizes is faster yet results in a higher-quality solution.
This suggests that ``live'' guidance, based on up-to-date observations, is essential for mitigating local congestion.

\subsection{Combined with Anytime Refinement}
In real-time planning situations, suboptimal MAPF methods can refine solutions within a given time budget after discovering initial solutions.
Here, there are two options:
\emph{(i)}~allocate more time to find better initial solutions and less time to refine them; or
\emph{(ii)}~allocate more time to refine them with quicker, but highly suboptimal initial solutions.
Since constructing guidance---either globally or locally---can improve the quality of the initial solution, albeit at a computational cost, we investigate which real-time planning strategy results in superior solutions at the end.
In particular, following~\cite{okumura2024lacam3}, we adopt a popular anytime MAPF scheme called LNS~\cite{li2021anytime,okumura2021iterative}, which repeatedly selects a subset of agents and refines their paths by PP~\cite{silver2005cooperative}.
For each subset selection, 1--30 agents are chosen at random, while four LNS processes are run concurrently~\cite{chan2024anytime}.

\Cref{fig:lns} compares LaCAM and those with guidance, followed by LNS refinement.
With fewer agents (i.e., less congestion), LNS quickly refines solutions, yielding similar final outcomes.
Meanwhile, densely populated scenarios show that superior initial solutions lead to substantial performance differences by the planning deadline, as smoothing highly interacting paths is challenging.
The results imply that, anytime MAPF methods with the current leading refinement scheme should devote more time to finding better initial solutions in such a dense setup.

We further compare LG+LNS with \texttt{lacam3}~\cite{okumura2024lacam3}, one of the most advanced MAPF solvers available today.
At its core, it employs GG and LNS, but it also uses other advanced techniques that are useful during the initial solution discovery and the refinement stages.
Note that the runtime required by \texttt{lacam3} for initial solutions is nearly identical to that of GG.
The result, shown in \cref{fig:lacam3}, reflects the trends observed thus far;
LG outperforms \texttt{lacam3} in finding better initial solutions, except on \mapname{warehouse}-type maps.
This leads to the final solutions that are comparable to and sometimes superior to those of \texttt{lacam3}.

\section{Conclusion}
This paper demonstrated that local guidance significantly improves the performance of the leading configuration-based MAPF method, LaCAM, incurring a non-negligible yet justifiable computational overhead.
The guidance construction is based on windowed decoupled planning, a concept that frequently appears in MAPF algorithms, and is therefore easy to implement.
Considering its ability to resolve congestion in bottleneck areas, where many agents inevitably use these, we believe it is worthwhile to invest more research effort in this concept, not limited to LaCAM.
One direct application is lifelong MAPF, where PIBT has proven to be a highly effective solution for building strong planners~\cite{jiang2025deploying,yukhnevich2025enhancing}.

\section*{Acknowledgments}
This research was supported by a gift from Murata Machinery, Ltd.

\bibliography{sty/ref-macro,ref}
\appendix
\onecolumn
\section*{Appendix}
\bigskip

\Cref{fig:heatmap2} presents further heatmaps corresponding to \cref{fig:heatmap}.
In \mapname{room-64-64-8}, LG vividly mitigates the local congestion especially in bottleneck regions where many agents need to use, resulting in significant cost reduction compared to GG.
Meanwhile, \mapname{warehouse-20-40-10-2-1} represents a situation in which GG is more effective, as discussed in the paper.

{
\setlength{\tabcolsep}{1pt}

% entry for room
\newcommand{\entryroom}[4]{
  \begin{scope}[xshift=0.22*#3\linewidth]
  {
    \node[anchor=south west] at (0, 0)
    {\includegraphics[width=0.22\linewidth]{fig/raw/heatmaps2/room-64-64-8_#2_visits_heatmap}};
    \node[] at (2.1, 4.55) {\small #1};
    \node[] at (2.1, 4.25) {\scriptsize cost improvement: #4};

    \node[anchor=north west] at (0.35, 0.1)
    {\includegraphics[width=0.205\linewidth,clip,trim={0.9cm 0.9cm 0 0}]{fig/raw/heatmaps2/room-64-64-8_#2_visits_histogram}};
    \node[anchor=north west,rotate=0] at (0.1, 0.2) {\tiny freq.};
    \node[anchor=north west] at (1.4, -0.92) {\tiny number of visits};
  }
  \end{scope}
}

% entry for warehouse
\newcommand{\entrywarehouse}[4]{
  \begin{scope}[xshift=0.22*#3\linewidth]
  {
    \node[anchor=south west] at (0, 0)
    {\includegraphics[width=0.22\linewidth]{fig/raw/heatmaps2/warehouse-20-40-10-2-1_#2_visits_heatmap}};
    \node[] at (2.1, 2.15) {\small #1};
    \node[] at (2.1, 1.85) {\scriptsize cost improvement: #4};

    \node[anchor=north west] at (0.35, 0.1)
    {\includegraphics[width=0.205\linewidth,clip,trim={0.9cm 0.9cm 0 0}]{fig/raw/heatmaps2/warehouse-20-40-10-2-1_#2_visits_histogram}};
    \node[anchor=north west,rotate=0] at (0.1, 0.2) {\tiny freq.};
    \node[anchor=north west] at (1.4, -0.92) {\tiny number of visits};
  }
  \end{scope}
}

% unified figure
\begin{figure*}[hp!]
\centering

% --- ROOM heatmaps ---
\begin{tikzpicture}
  \entryroom{LaCAM}{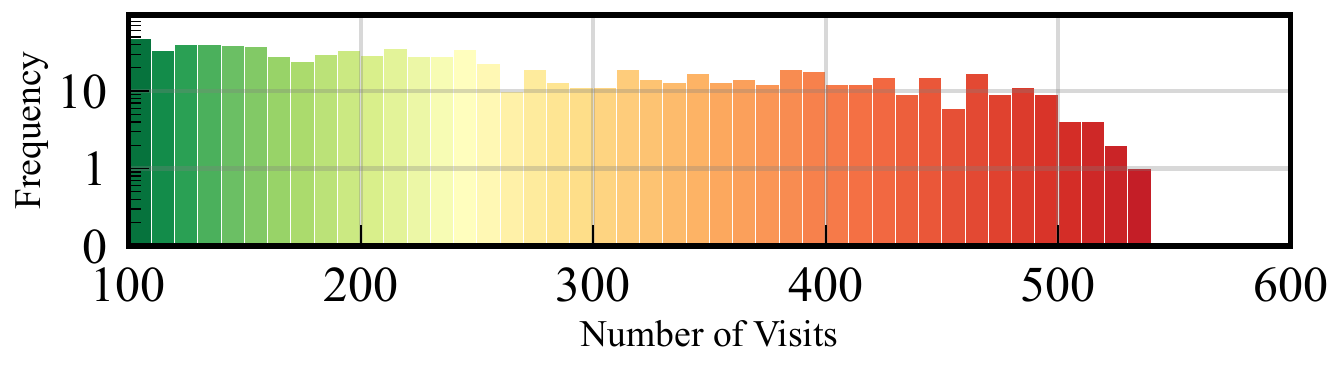}{0}{0.0\%} % 284209  -> tikz math cannot calculate this...
  \entryroom{GG}{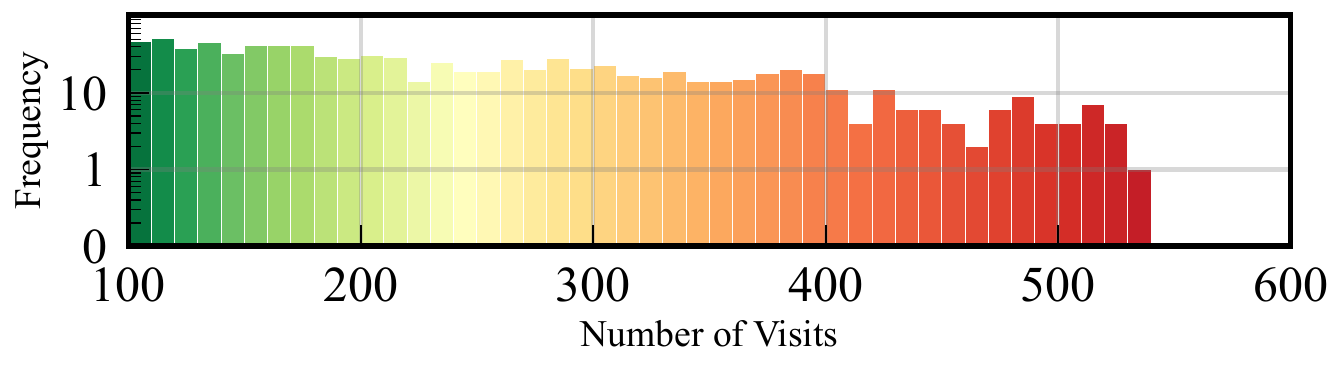}{1}{3.1\%} % (284209 - 338314) / 284209
  \entryroom{LG}{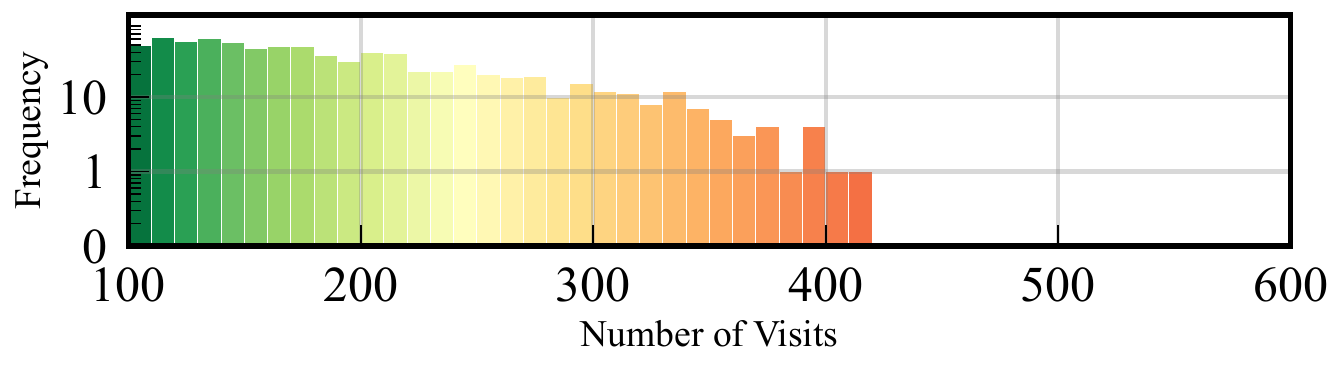}{2}{31.5\%} % (284209 - 239196) / 284209
  \entryroom{GG+LG}{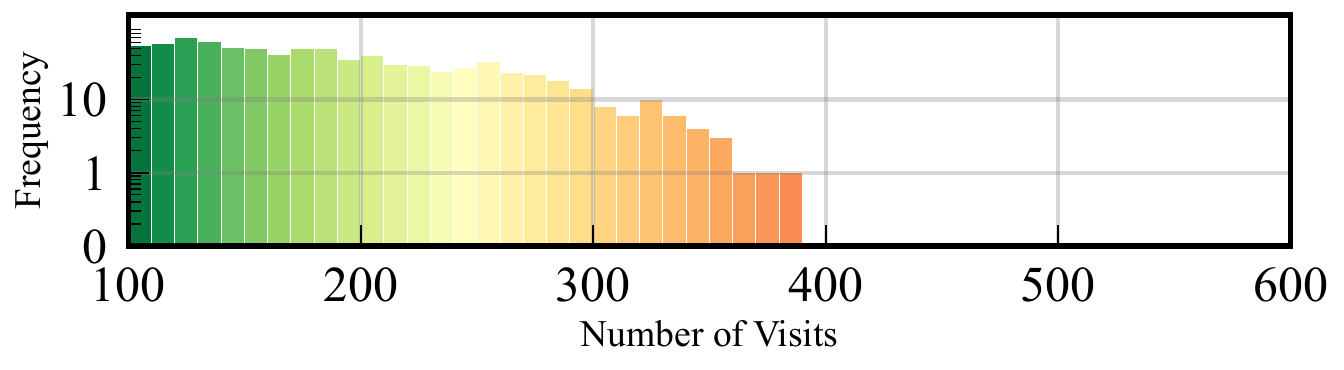}{3}{30.8\%} % (284209 - 241655) / 284209
  \node[anchor=south west] at (0.89\linewidth, -0.05)
  {\includegraphics[width=0.05\linewidth,clip,trim={0 0 0.8cm 0}]{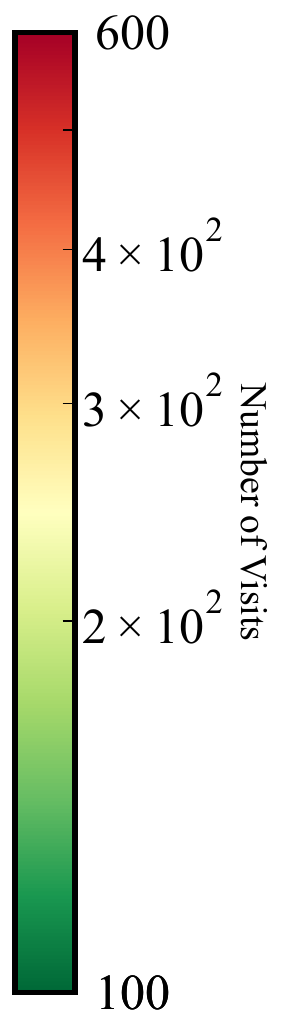}};
  \node[anchor=south east,rotate=-90] at (0.95\linewidth, 1.0) {\small number of visits};
\end{tikzpicture}
\smallskip

% --- WAREHOUSE heatmaps ---
\begin{tikzpicture}
  \entrywarehouse{LaCAM}{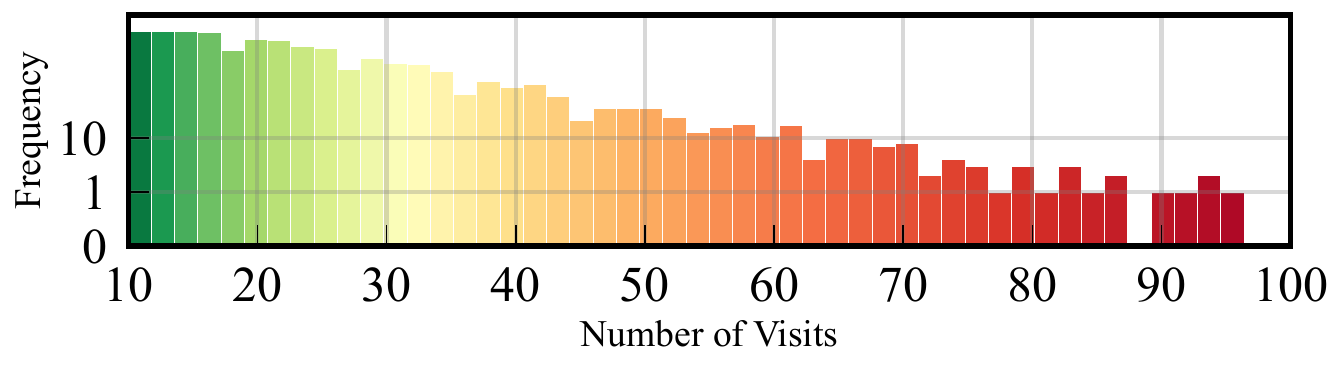}{0}{0.0\%} % 284209  -> tikz math cannot calculate this...
  \entrywarehouse{GG}{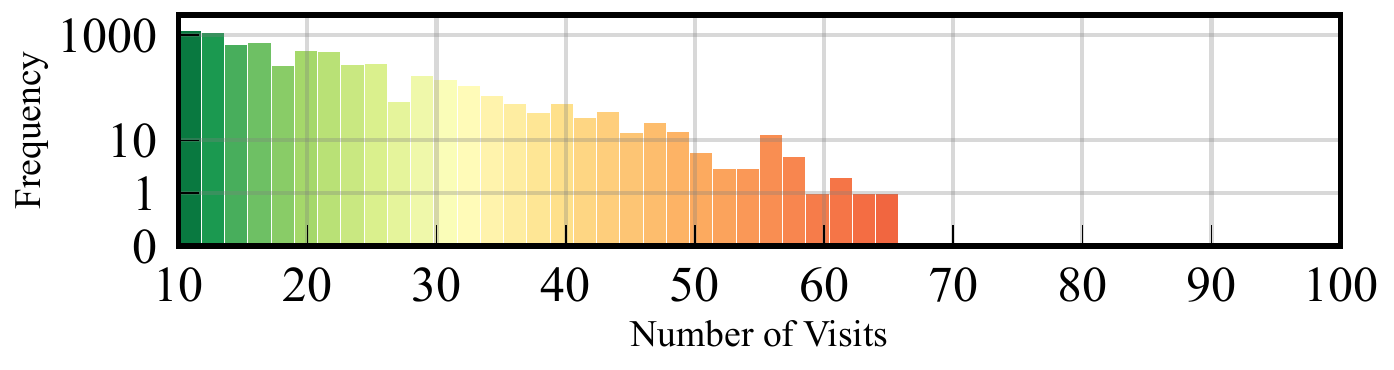}{1}{33.3\%} % (284209 - 189316) / 284209
  \entrywarehouse{LG}{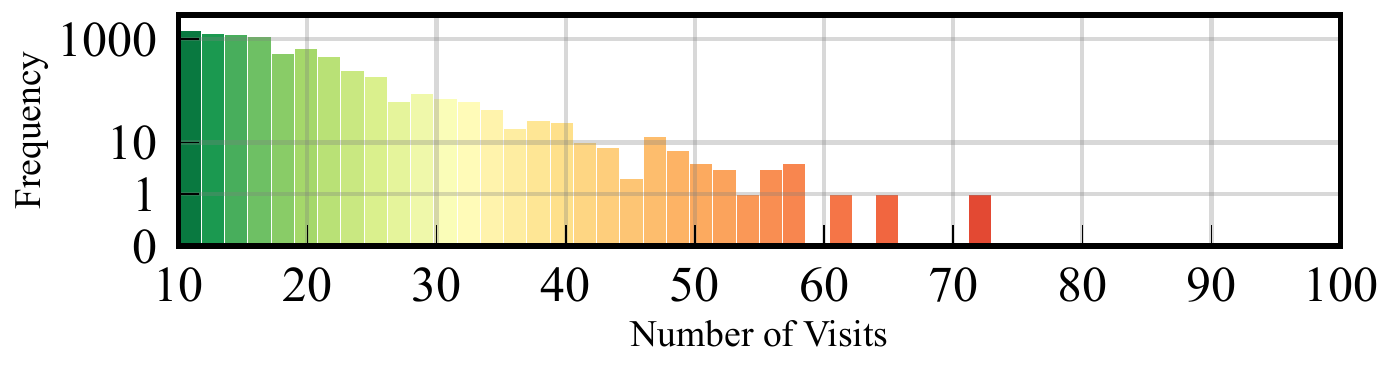}{2}{26.2\%} % (284209 - 209517) / 284209
  \entrywarehouse{GG+LG}{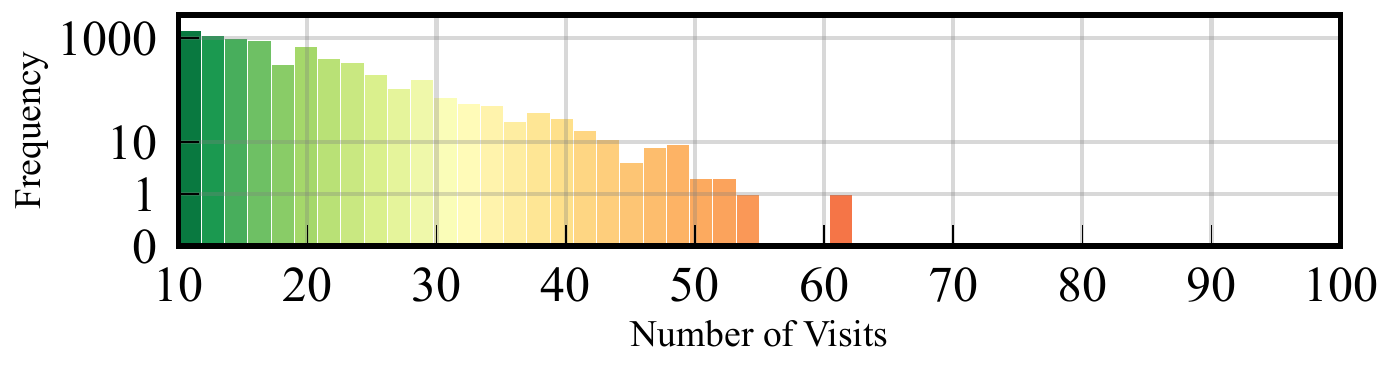}{3}{31.9\%} % (284209 - 193531) / 284209
  \node[anchor=south west] at (0.89\linewidth, -1.0)
  {\includegraphics[width=0.033\linewidth,clip,trim={0 0 0.8cm 0}]{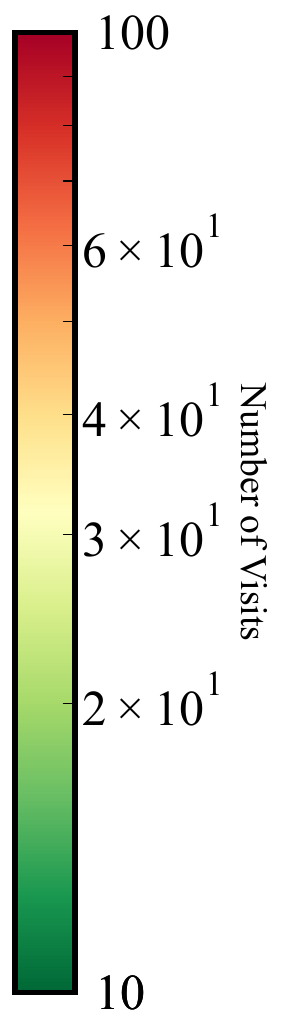}};
  \node[anchor=south east,rotate=-90] at (0.95\linewidth, -0.8) {\small number of visits};
\end{tikzpicture}

\caption{
\emph{Top:}~Heatmaps showing how many times each vertex is used in each solution on \mapname{room-64-64-8} with $1,000$ agents.
\emph{Bottom:}~For \emph{warehouse-20-40-10-2-1} with $1,000$ agents.
}
\label{fig:heatmap2}
\end{figure*}
}

\end{document}